\documentclass[smallextended]{svjour3}       % onecolumn (second format)
\smartqed  % flush right qed marks, e.g. at end of proof
\usepackage{graphicx}
\usepackage{natbib}
\usepackage{amsmath}
\usepackage{url}
\usepackage{amssymb}

\usepackage{mathptmx}      % use Times fonts if available on your TeX system
\begin{document}

   \title{Monitoring fast solar chromospheric activity: the MeteoSpace project}

   \titlerunning{The MeteoSpace project}

   \authorrunning{J.-M. Malherbe et al.}

   \author{Jean-Marie Malherbe
          \and
          Thierry Corbard
          \and
          Ga\"{e}le Barbary
           \and
          Fr\'{e}d\'{e}ric Morand
           \and
          Claude Collin
           \and
          Daniel Crussaire
           \and
          Florence Guitton
          }

   \institute{J.-M. Malherbe \at
              LESIA, Observatoire de Paris, PSL Research University, CNRS, Meudon,
              France\\
              \email{Jean-Marie.Malherbe@obspm.fr}
         \and
             T. Corbard \at
             Lagrange, Observatoire et Universit\'{e} de la C\^{o}te d'Azur, CNRS, Nice,
             France\\
             \email{Thierry.Corbard@oca.eu}
         \and
              G. Barbary \at
              LESIA, Observatoire de Paris, PSL Research University, CNRS, Meudon,
              France\\
              \email{Gaele.Barbary@obspm.fr}
         \and
              F. Morand \at
              Calern station, Observatoire et Universit\'{e} de la C\^{o}te d'Azur, CNRS, Caussols,
             France\\
              \email{Frederic.Morand@oca.eu}
         \and
              C. Collin \at
              LESIA, Observatoire de Paris, PSL Research University, CNRS, Meudon,
              France\\
              \email{Claude.Collin@obspm.fr}
         \and
              D. Crussaire \at
             LESIA, Observatoire de Paris, PSL Research University, CNRS, Meudon,
              France\\
               \email{Daniel.Crussaire@obspm.fr}
         \and
              F. Guitton \at
              Lagrange, Observatoire et Universit\'{e} de la C\^{o}te d'Azur, CNRS, Nice,
             France\\
              \email{florence.Guitton@oca.eu}
               }

\date{Received: 15 December 2021 / Accepted: date}

\maketitle

\begin{abstract}
We present in this reference paper an instrumental project dedicated
to the monitoring of solar activity during solar cycle 25. It
concerns the survey of fast evolving chromospheric events implied in
Space Weather, such as flares, coronal mass ejections, filament
instabilities and Moreton waves. Coronal waves are produced by large
flares around the solar maximum and propagate with chromospheric
counterparts; they are rare, faint, difficult to observe, and for
that reason, challenging. They require systematic observations with
automatic, fast and multi-channel optical instruments. MeteoSpace is
a high cadence telescope assembly specially designed for that
purpose. The large amount of data will be freely available to the
solar community. We describe in details the optical design, the
qualification tests and capabilities of the telescopes, and show how
waves can be detected. MeteoSpace will be installed at Calern
observatory (C\^{o}te d'Azur, 1270 m) and will be in full operation
in 2023.

\end{abstract}

%%
%%________________________________________________________________

\section{Introduction} \label{sec:Intro}

Solar activity is the primary driver of Space Weather phenomena,
such as Coronal Mass Ejections (CMEs), flares, energetic particles
and wind streams which may reach and perturb the Earth environment.
The solar perspectives of Space Weather are reviewed by
\cite{Schwenn2006} and more recently by \cite{Temmer2021}, while the
terrestrial consequences are discussed by \cite{Pulkkinen2007}.
Solar activity occurs over timescales ranging from seconds to
minutes in flares and CMEs, which are fast evolving and highly
dynamic events, originating in the solar atmosphere from
non-potential magnetic energy. It can be stored in bright faculae (B
$\approx$ 100 G, 1 G = 10$^{-4}$ T), sunspots (B $\approx$ 1000 G)
and large coronal loops above active regions (review by
\cite{Reale2010}). The monitoring of solar activity requires a
systematic survey of the chromosphere (8000 K), as it is the source
of solar events, which propagate to the hot corona above (1-2 MK or
more). For that purpose, ground-based networks with many stations
around the world have been organized, such as the Global H$\alpha$
network (GHN, \cite{Steinegger2000}), the Continuous H$\alpha$
Imaging Network (CHAIN, \cite{Ueno2010}) or the GONG H$\alpha$
network with seven identical telescopes \citep{Harvey2011}. In
space, observations of various EUV emission lines, formed in the
temperature range 0.1-10 MK of the corona, started with the
SOHO/ESA/NASA mission (1996) and continue with the improved Solar
Dynamics Observatory (SDO/NASA, 2010). New data are now available
from recent spacecrafts, such as Parker Solar Probe (NASA, 2018) and
Solar Orbiter (ESA, 2020).

The Space Weather is mainly affected by CMEs \citep{Chen2011} and
energetic flares \citep{Oloketuyi2019}. The integrated X-ray flux of
the Sun is permanently monitored by the GOES/NASA satellites and is
a good indicator of solar activity and the flare energy. The flux is
usually symbolized by letters A, B, C, M and X, respectively for
10$^{-8}$ to 10$^{-4}$ W m$^{-2}$ in logarithmic scale, and a
multiplicative number. Only C, M and X-class flares are significant.
Among them, energetic events (M and X) are relatively rare
(respectively 9\% and 1\%, with only 50 flares between X2.6 and X28
during the past 25 years). The largest recent event (4 November
2003, X28, about $10^{25}$ J) could be compared to the famous and
historic Carrington flare (1 September 1859 seen in white light,
maybe X50). Highest energy flares (X-Class) trigger fast coronal MHD
shock waves \citep{Warmuth2015} which propagate at
super-magnetosonic speeds (500-1000 km s$^{-1}$) and can produce
chromospheric counterparts, discovered much earlier by
\cite{Moreton1960}. They are often described as the signature of the
downward compression by the moving coronal shock above.
Chromospheric Moreton waves are fast and rare phenomena, so that
study cases are not very numerous (\cite{Zhang2001},
\cite{Narukage2004}, \cite{Balasubramaniam2007}, \cite{Zhang2011},
\cite{Asai2012}, \cite{Krause2015}, \cite{Admiranto2015},
\cite{Krause2018}, \cite{Cabezas2019}). The usual observing cadence
is about 45-60 s, but faster frame rates (5 images/minute or more)
would be better to investigate with more details the chronology and
the physics of such events.

The MeteoSpace (MTSP) instrument presented in this paper will
provide a unique opportunity to monitor highly dynamic phenomena
around the next solar maximum (2025). Our paper is organized as
follows. Section ~\ref{sec:MTSP} summarizes the architecture of the
MTSP project, while Section ~\ref{sec:cak} presents the CaII K
channel. The goals and capabilities of the two H$\alpha$ telescopes
are described in Section ~\ref{sec:ha}, while Section
~\ref{sec:moreton} discusses the detection of Moreton waves, which
is the most challenging purpose of the H$\alpha$ channels. At last,
Section ~\ref{sec:exten} describes a possible extension for the
future.

\section{Architecture of the MeteoSpace (MTSP) project} \label{sec:MTSP}

MTSP has an historical background. The chromosphere is observed
daily at Meudon with Deslandres's spectroheliograph since 1908. It
produces now (x, y, $\lambda$) data-cubes from which monochromatic
images are derived \citep{Malherbe1}. In 1957, during the
International Geophysical Year (IGY), the french astronomers decided
to organize a survey of H$\alpha$ chromospheric activity at high
cadence (60 s) using the Lyot filter technology \citep{Lyot}. Two
instruments were built for Meudon and Haute Provence observatories
(\cite{Grenat}, \cite{Michard}). In 1965, a new generation of
tunable Lyot filters allowed to observe sequentially the H$\alpha$
line centre and wings, in order to measure Doppler-shifts. This
routine, improved in 1985, was stopped in 2004, due to the filter
obsolescence and the retirement of operators. About 7 million images
have been recorded. In the competitive context of Space Weather
research and applications, we wish to restart a survey with new
automatic telescopes and faster cadence (10-15 s), on a sunnier site
in order to increase the probability of catching rare events, such
as large X-class flares and Moreton waves. MTSP is a joint project
between Paris (OP) and C\^{o}te d'Azur (OCA) observatories. Overall
characteristics are summarized in \cite{Malherbe2}. We provide here
much more instrumental details in order to constitute a reference
paper for MTSP.

\begin{figure}
   \centering
   \includegraphics[width=\textwidth]{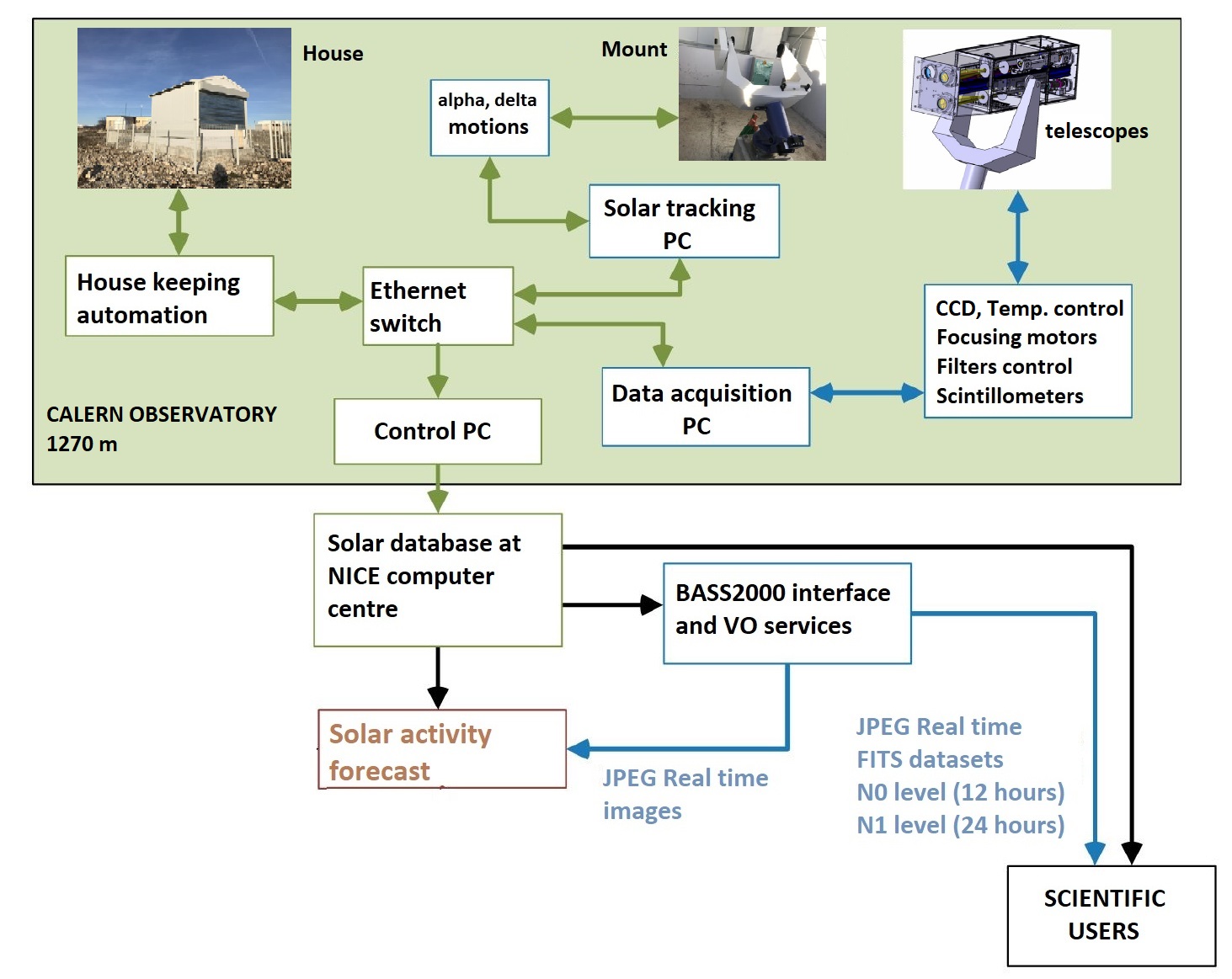}
      \caption{The MTSP project architecture. The green box is located at Calern
      station, and the database at Nice observatory. Observations will also be available through
      the BASS2000 data centre with additional Virtual Observatory (VO) services. Real time
      images (JPEG)
      together with FITS scientific data-sets will be freely delivered to the international community.}
         \label{MTSP}
   \end{figure}

The architecture of the MTSP project (Figure~\ref{MTSP}) is
organized around three main tasks:

\begin{itemize}
  \item Automation procedures (OCA)
  \item The real time data processing and archiving procedures (OCA)
  \item The telescopes and data acquisition system (OP)
\end{itemize}

The assembly of three telescopes (two H$\alpha$ 6563 \AA~ channels
for solar activity, and a CaII K 3934 \AA~ magnetic proxy) is built
by OP. A new housing and equatorial mount have been installed at
Calern observatory (1270 m, OCA, Figure~\ref{House}). The
instruments (Figure~\ref{Telescopes}) will be integrated mid 2022,
then tested and commissioned in 2023 for systematic and automatic
observations. Data will be freely available to the international
community, without any delay, both for operational and scientific
purpose. MSTP comes with a specific database (located at Nice
computer centre with 100 TB of disk storage), with possible access
through the BASS2000 solar database, and will offer later virtual
observatory services.

\subsection{Automation procedures}

\begin{figure}
   \centering
   \includegraphics[width=\textwidth]{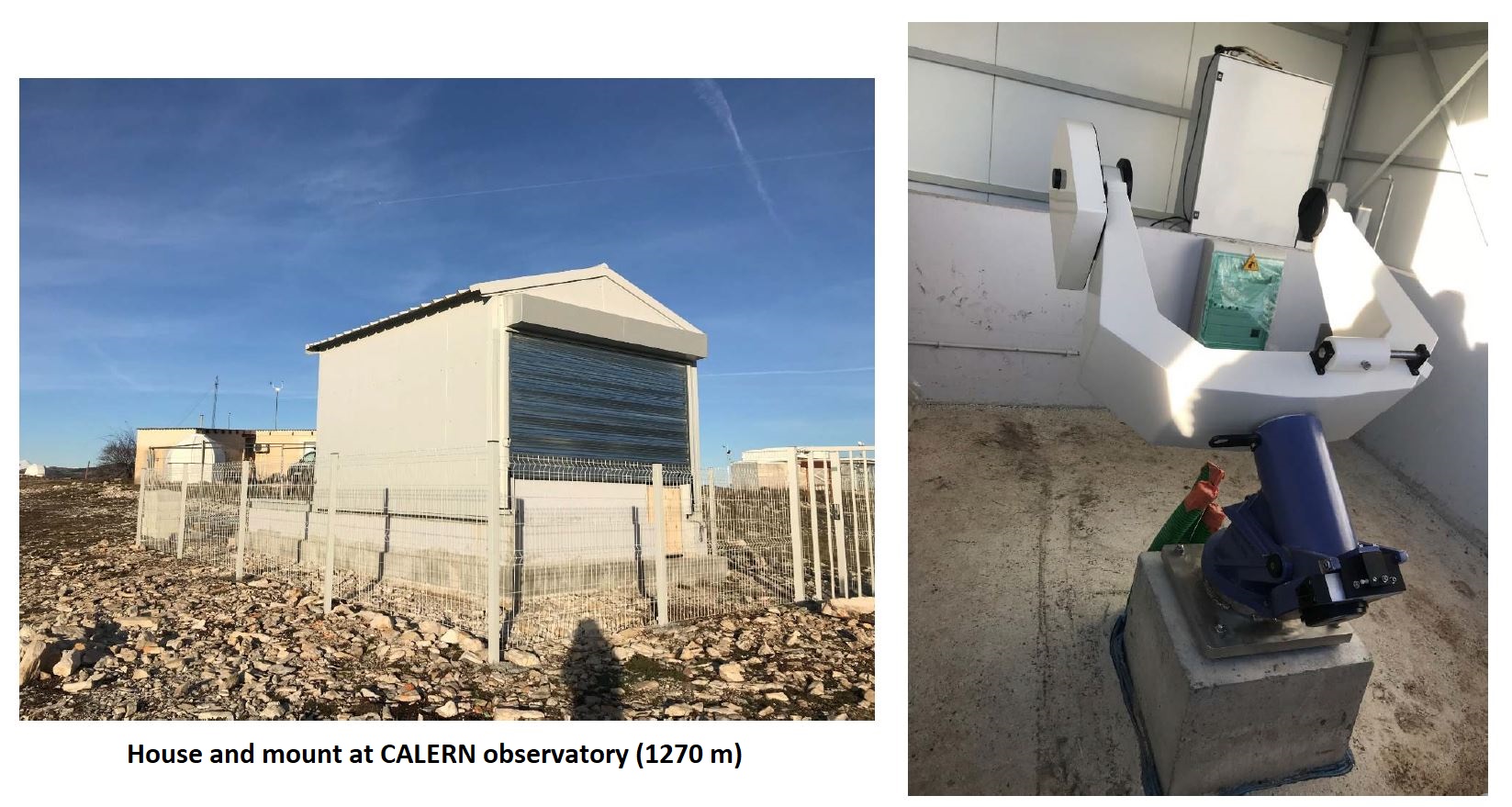}
      \caption{The MTSP housing. Left: the MTSP building at Calern observatory (1270 m, C\^{o}te d'Azur). It
      moves on two rails and is computer controlled. It is closed by metallic curtains and connected
      to many environment detectors (such as cameras and meteorological devices) for automatic operation.
      Right: the equatorial mount waiting for the instruments.}
         \label{House}
   \end{figure}

MTSP is the first automatic solar telescope assembly available in
France (other solar instruments are under human control). The
telescopes (Figure~\ref{Telescopes}) and housing
(Figure~\ref{House}) are under the supervision of a control/command
computer (Figure~\ref{MTSP}). This PC is interconnected with other
slave machines, such as the data acquisition computer (which handles
cameras, filters, motor focus, scintillometers, thermal units), the
solar tracking computer (which drives the equatorial mount) and the
House Keeping (HK) automation. HK is a complex system for a fully
automatic instrument, because many environment detectors and cameras
must be controlled permanently, concerning the telescope status and
meteorology (clouds, temperature, humidity, wind). In particular,
the HK system will decide to start or interrupt observations
according to informations provided by the environment parameters.
Under good observing conditions, the HK will roll the housing and
open front curtains; the tracking PC will catch and follow the Sun,
and the instrument PC will start data acquisition after checking the
different components. In case of danger (such as wind, rain, snow),
the HK procedure will return an alarm to the central PC which will
interrupt observations, order the mount to return to the garage
position and ask the HK to close the building. SMS messages will
inform the technical staff of Calern observatory, in particular in
case of alarms.

\subsection{Real time data processing and archiving procedures}

Raw 12 bits data are collected by the acquisition PC and transmitted
to the control computer. There are two data levels, N0 and N1. N0
are raw data written in FITS format. For that purpose, FITS keywords
with exhaustive data descriptors are simply added and N0 images are
directed to the Nice archive via the fast network link between the
observing site and the data centre. N1 is the first level of
scientific data. For each image, the Sun radius and centre are
detected and the solar disk is centred in the 35 arcmin FOV. A
rotation is then applied in order to present the North up. The dark
current is subtracted. Corrections of optical distortions are
applied, if needed. A 8 bits JPEG real time image is derived and is
immediately uploaded into the database. Keywords are added to
scientific images to form FITS data-sets, which will be transferred
after observations (by night), so that the delay for scientific data
will be about 12 hours. The query system of the Nice database will
offer, for each channel, individual images or large data-sets (zip
or tar.gz files) that will be built on line using the time interval
provided by the requestor. The database is dimensioned to store all
observations of cycle 25 and will start to operate in 2023. In
parallel, the query system will be incorporated to the solar
BASS2000 data portal. Progressively, it will offer virtual
observatory services, so that links to other wavelengths will be
offered together with MTSP data (such as EUV from SDO/AIA or radio
from Nan\c{c}ay Radioheliograh). Observations will start with two
operating modes: a standard one in the case of low activity level
(60 s cadence), and a fast mode (10-15 s) in the case of high level.
This mode will be systematically used at the approach of the next
solar maximum (2025), or will be triggered by indicators such as the
real-time GOES/NASA X-ray flux.

\subsection{The telescopes}

\begin{figure}
   \centering
   \includegraphics[width=\textwidth]{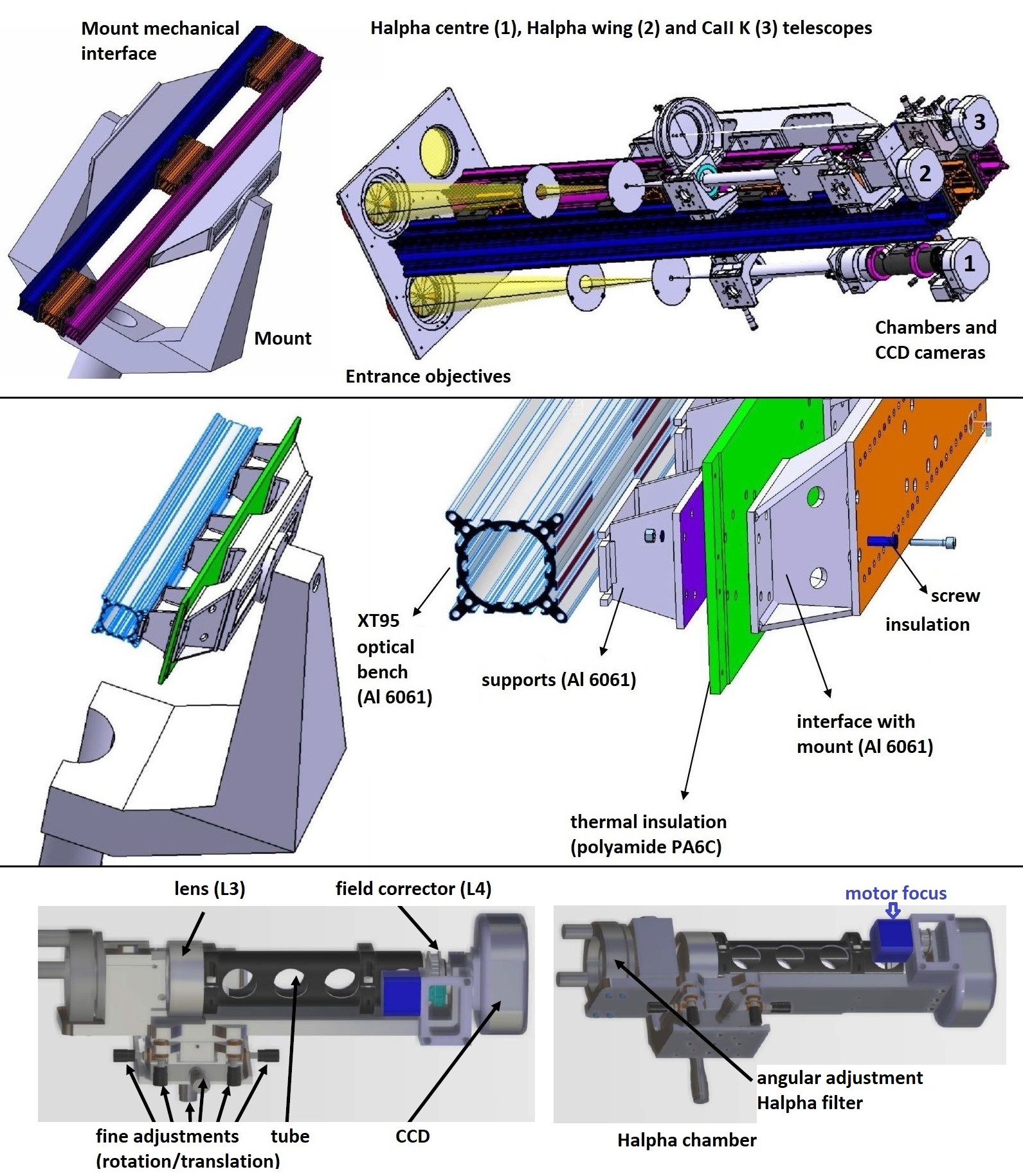}
      \caption{The MTSP instruments. Top: the mechanical structure and the three telescopes (1, 2 for H$\alpha$ and 3 for CaII K). There is room for
      a fourth telescope, if needed in the future. The instruments
      are mounted on two XT95 optical benches (aluminium 6061) and
      are enclosed in a thermally
      regulated and isolated box (not shown). The volume is 0.5 $\times$ 0.5 $\times$ 1.80 m$^{3}$ and
      the mass is about 120 kg. Centre: details of the complex mechanical interface between the equatorial mount
      and the two optical benches. Bottom: details of the H$\alpha$ chambers (see also Figure~\ref{lunetteha} of Section ~\ref{sec:ha}).}
         \label{Telescopes}
   \end{figure}

Three telescopes, with equivalent focal length 983 mm, forming a
solar image of 9.14 mm diameter, have been developed
(Figure~\ref{Telescopes}). The CaII K line channel provides a
magnetic proxy. It reveals the magnetized areas such as dark spots
(1000 G typical) and bright facular regions (100 G typical), through
broad-band interference filters integrating the line core (K3), K2
close wings ($\pm$0.2 \AA) and partially K1 far wings ($\pm$1.5
\AA). The two H$\alpha$ channels monitor chromospheric activity,
such as filament instabilities, flares, CMEs and eventually Moreton
waves. They use narrow bandpass filters. In order to have a compact
instrument, we selected Fabry-P\'{e}rot filters instead of Lyot
filters previously used at Meudon.

\begin{figure}
   \centering
   \includegraphics[width=\textwidth]{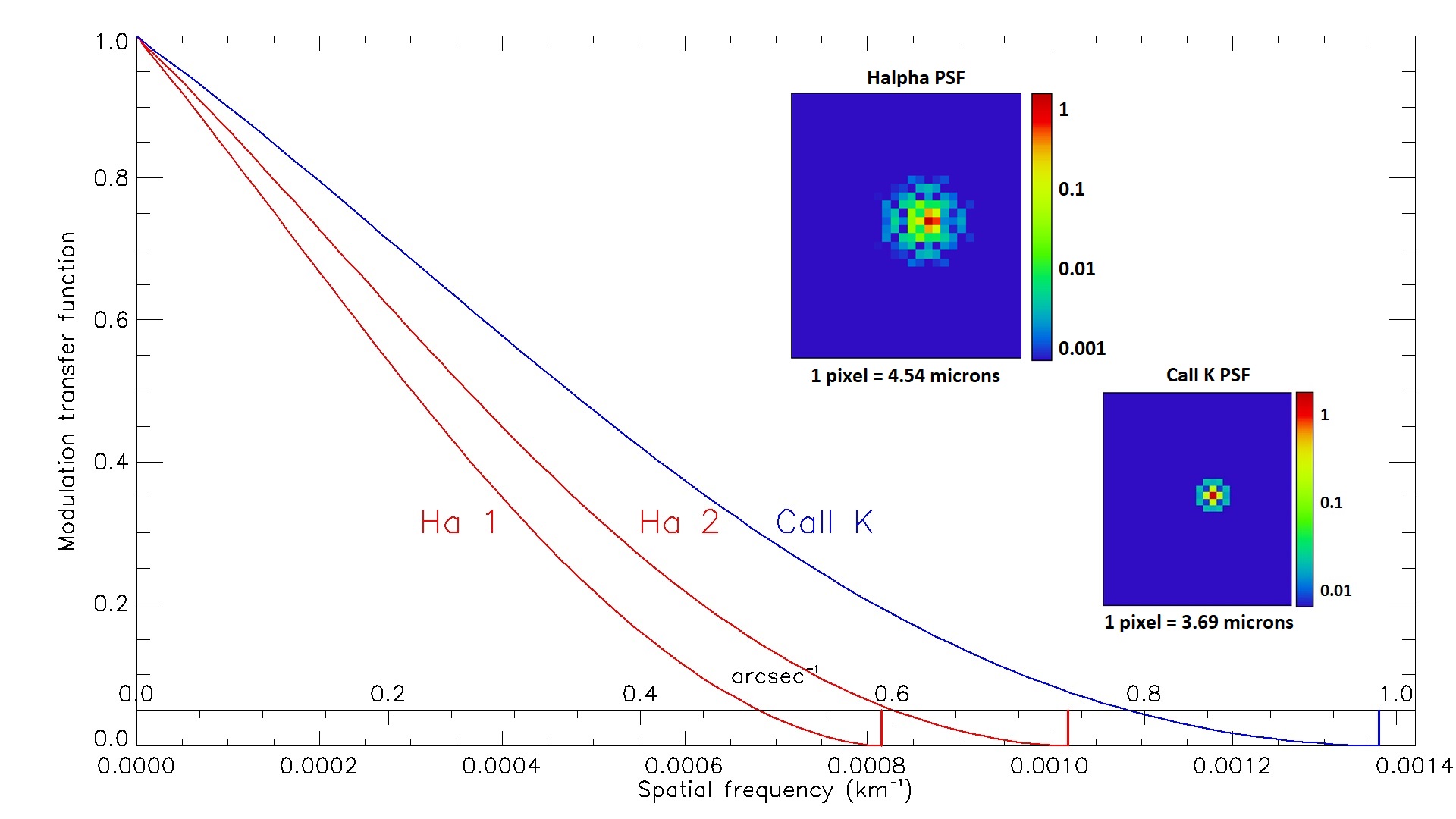}
      \caption{The theoretical MTF of the H$\alpha$ (red) and CaII K (blue)
      telescopes.
      Abscissa: spatial fequency in km$^{-1}$ or in arcsec$^{-1}$. Cutoff frequencies
      are indicated (bars). The computed PSFs, in the detector plane, are reported (the pixels
      correspond to the CCD sampling). The H$\alpha$ PSF
      is for the border of the FOV, while the CaII K PSF is for the centre.}
         \label{mtf}
   \end{figure}

The telescopes (except filters) have been simulated with the Zemax
software. Figure~\ref{mtf} shows the Modulation Transfer Functions
(MTF) of the aperture, including the CCD sampling effect. The cutoff
frequencies are 1.36, 0.815 and 1.02 10$^{-3}$ km$^{-1}$,
respectively for the CaII K (80 mm aperture) and the two H$\alpha$
channels (80 mm and 100 mm aperture). The Point Spread Functions
(PSF) in the image plane are reported. The CaII K and H$\alpha$
telescopes exhibit Strehl ratios of 88\% and 93\% respectively, and
the ensquared energy of a point source is 70\% in a box of only 2
$\times$ 2 pixels for all instruments. There is almost no optical
distortion. The RMS radius of spot diagrams varies from 32\% to 65\%
of the Airy spot radius from the FOV centre to the border, in the
case of H$\alpha$ telescopes.

\begin{figure}
   \centering
   \includegraphics[width=\textwidth]{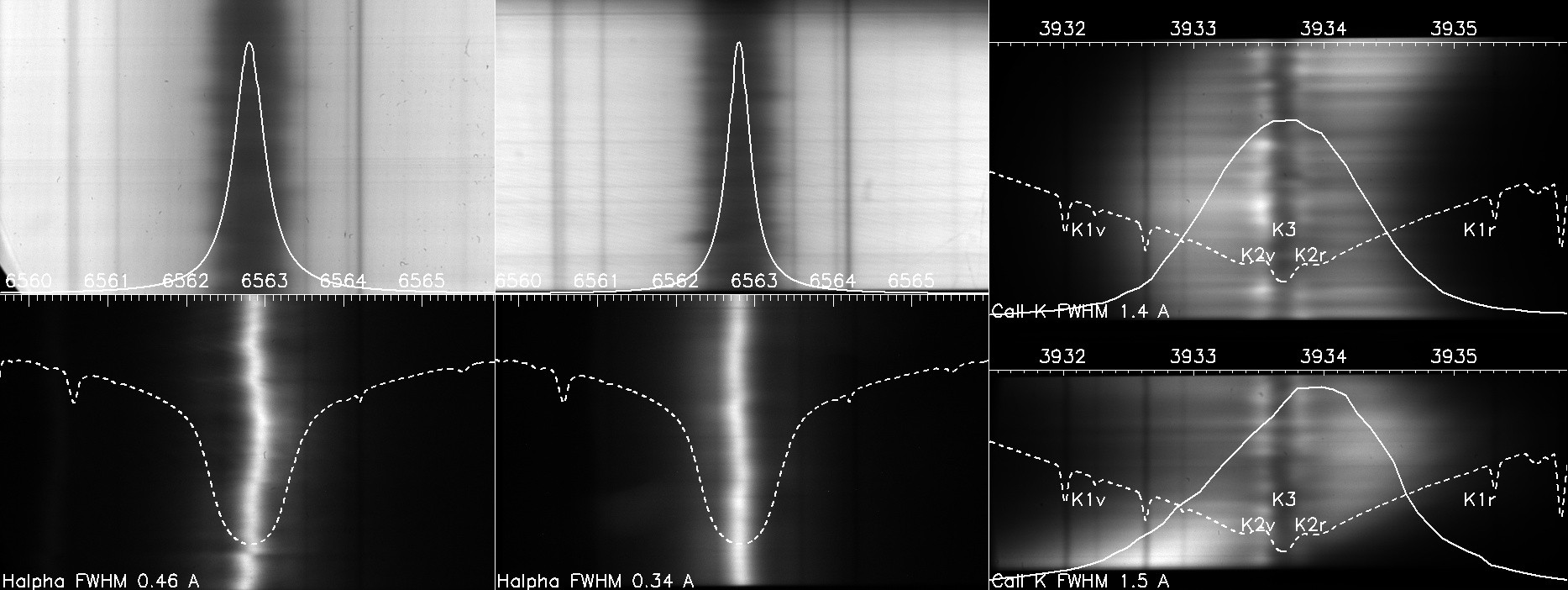}
      \caption{The MTSP Filter bandwidths. Filters have been controlled with
      the high dispersion spectrograph of Meudon Solar Tower. Left: H$\alpha$ line without (top)
      or with (bottom) filter 1. The pupil integrated transmittance (solid line, top) over the surface filter is
      drawn (0.46 \AA~ FWHM). Middle: H$\alpha$ line without (top)
      or with (bottom) filter 2. The pupil transmittance is 0.34 \AA~ FWHM. Right: CaII K
      line as seen
      through the two available filters of respectively 1.5 (filter 1, bottom) and 1.4 \AA~ FWHM (spare filter 2, top). The transmittance
      (solid line) includes K2v, K3 and K2r features, and partially K1v and K1r (v, r letters meaning respectively violet and red wings).}
         \label{Filters}
   \end{figure}

The optical design and filter tests are detailed in sections
~\ref{sec:cak} and ~\ref{sec:ha}, respectively for CaII K and
H$\alpha$ telescopes. However, Figure~\ref{Filters} summarizes
briefly the overall capabilities of MTSP devices. We have two
H$\alpha$ Fabry-P\'{e}rot filters, in pupil plane, with respectively
0.46 \AA~ FWHM (for line wings, filter 1) and 0.34 \AA~ FWHM (for
line centre, filter 2). We also have two interference CaII K
filters, in image plane, with respectively 1.5 and 1.4 \AA~ FWHM
(the second one is a spare). The instruments are enclosed inside a
temperature regulated box at 30$\pm$1$^{\circ}$C by active heating
and passive cooling, but Fabry-P\'{e}rot filters have their own and
high precision heating oven (typical operating values of 65 and
40$^{\circ}$C respectively for filters 1 and 2). CCDs are isolated
with their own Peltier and air cooling systems; according to the
readout noise at 8 MHz pixel rate, their dynamic range is about 2700
(12 bits) for a typical signal to noise ratio of 100. Exposure times
are a few ms only.

The H$\alpha$ telescopes, dedicated to fast solar events, can
observe the chromosphere (8000 K) at high cadence (10-15 s), this is
4 times faster than GONG observations or previous Meudon routines,
and 3 times faster than SDO/AIA in EUV emission lines of the hot
corona. The pixel size is about 1 arcsec (2 times better than the
previous routines at Meudon); it is difficult to achieve a better
resolution with small ground based instruments observing the Sun.

\section{The CaII K channel} \label{sec:cak}

\begin{figure}
   \centering
   \includegraphics[width=\textwidth]{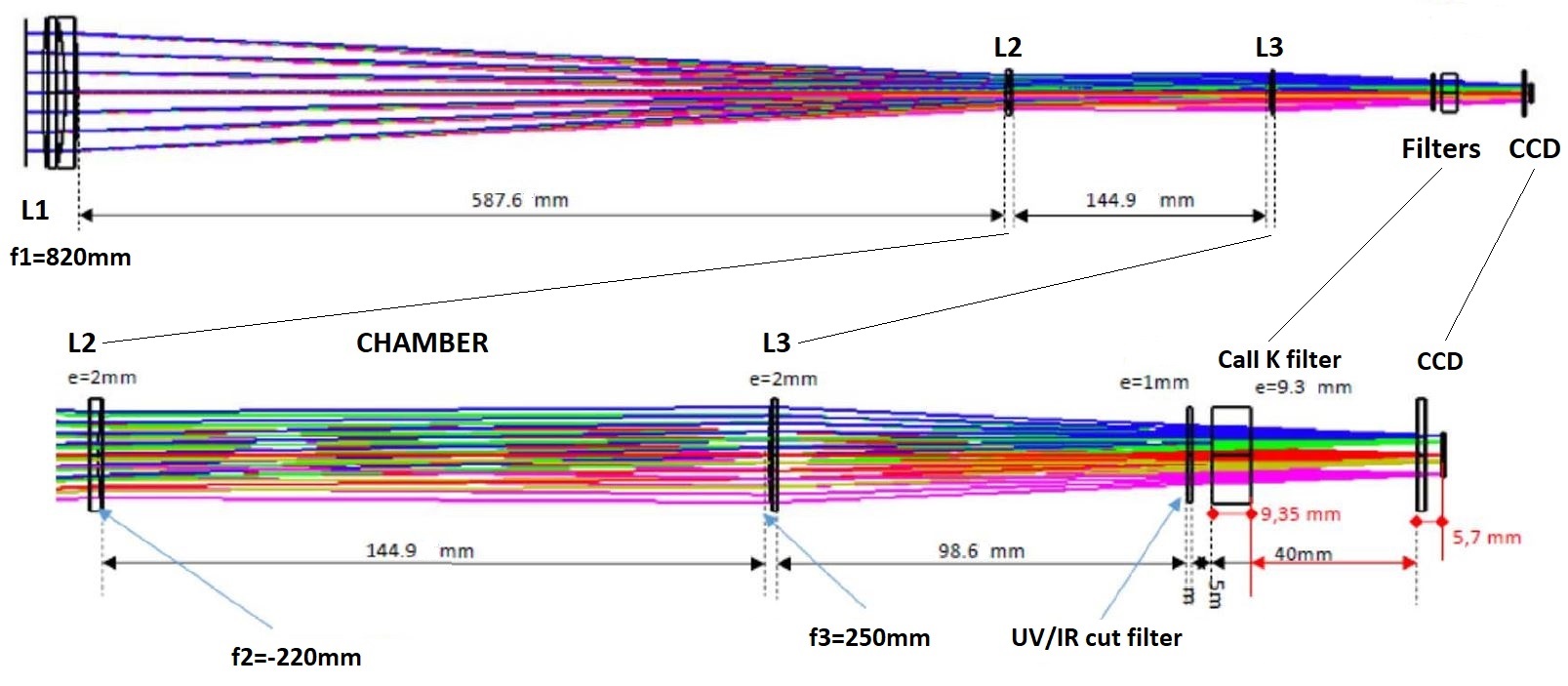}
      \caption{The MTSP CaII K channel. Top: the overall design of the telescope. Bottom: details
      of the chamber. L2/L3 increase the focal length from 820 to 983 mm. Filters are in the image plane at F/12.3.
      The chamber has a motor focus.}
         \label{lunetteca}
   \end{figure}

The CaII K telescope (Figure~\ref{lunetteca}) has an aperture of 80
mm. It is composed of a Takahashi FS102 objective (820 mm focal
length) followed by a magnifying chamber ($\times$ 1.2), a Barr
Associates 1.5 \AA~ FWHM filter (in image plane) and an interline
cooled CCD camera (QSI 690, pixel size = 3.69 $\mu$m = 0.78 arcsec).
The system is protected by an UV/IR cutoff filter. The Airy spot
size is 5.9 $\mu$m or 1.24 arcsec resolution.

\begin{figure}
   \centering
   \includegraphics[width=\textwidth]{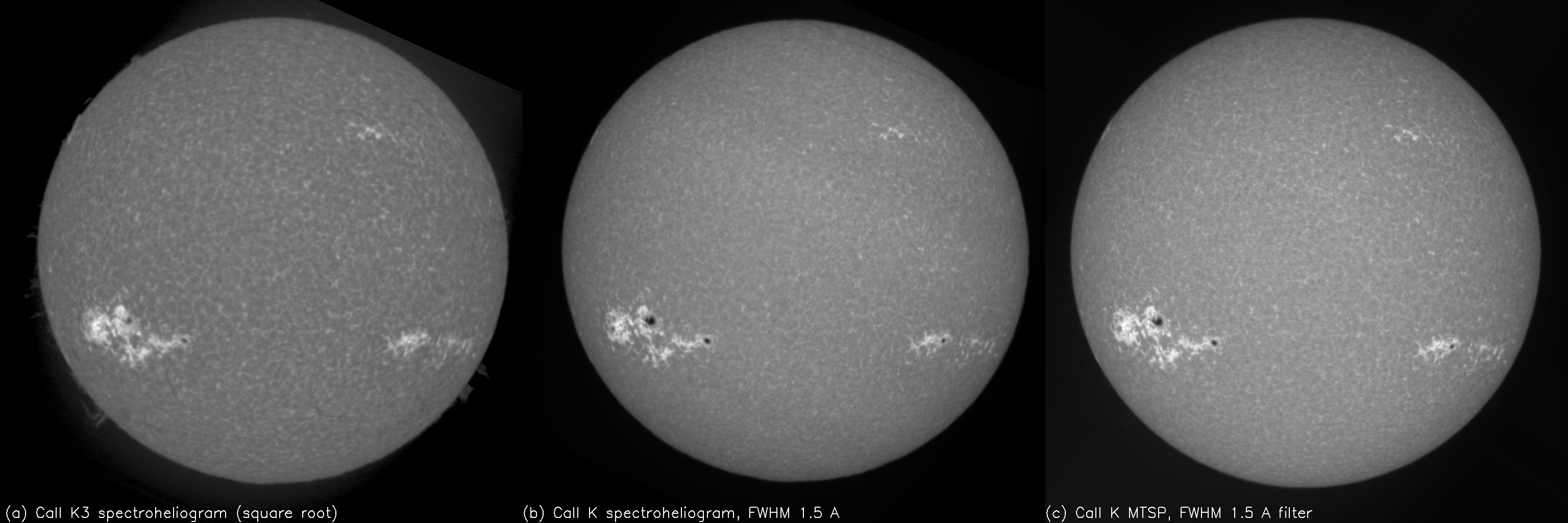}
      \caption{Comparison of Meudon CaII K3 spectroheliograms and MTSP CaII K images, 26 November 2020.
      (a) K3 spectroheliogram (30 s surface scan by the slit, square root to decrease the dynamics). (b) the spectroheliograph
      (x, y, $\lambda$) data-cube integrated over the MTSP wavelength transmission curve. (c) the MSTP K image, 1.5 \AA~ FWHM (2 ms exposure time,
      the waveband covers K2v, K3 and K2r central features of the CaII K line, and partially K1v and K1r wings).}
         \label{cakimages}
   \end{figure}

The interference filters were tested with the powerful 14 m
spectrograph of Meudon Solar Tower at F/75 (R = 300000, 6 m\AA~
spectral line sampling in order 15). We found that the central
wavelength (CWL) is temperature (T) dependant, according to the law
$\Delta\lambda_{CWL}$ = C $\Delta$T, where C = 0.07 \AA/$^{\circ}$C
is the measured temperature coefficient. It is also function of the
incidence angle $\theta$, and follows the law $\Delta\lambda_{CWL}$
= K $\theta^{2}$ with K = -0.15 \AA/degree$^{2}$. As the theory
gives K = - $\frac{1}{2}$ $\frac{\lambda_{0}}{n^{2}}$ ($\theta$ in
radians), where n is the effective index of refraction and
$\lambda_{0}$ the line wavelength, we derived from the measurements
n $\approx$ 2.0. With an aperture of F/12.3 (half cone of
2.3$^{\circ}$), the numeric integration over the light cone shows
that the CWL is blue-shifted (-0.41 \AA), but this is almost
compensated (+0.49 \AA) by the operating temperature (+7$^{\circ}$C
above the specification of the manufacturer), so that the resulting
CWL shift is small in comparison to the FWHM (the transmittance of
the filter centre is displayed in Figure~\ref{Filters}, right). This
figure also reveals that the surface filter is not uniform. We found
that that the CWL varies from the centre to the border according to
the law $\Delta\lambda_{CWL}$ = $\kappa$ x$^{2}$, where x is the
distance to the filter centre. We measured $\kappa$ = -0.006
\AA/mm$^{2}$ and +0.003 \AA/mm$^{2}$ respectively for filters 1 and
2 (the spare filter). It means that the CWL varies between the
centre of the solar disk and the limb (0 $\leq$ x $\leq$ 4.6 mm),
numerically $\Delta\lambda_{CWL}$ = -0.13 \AA~ or +0.07 \AA~,
respectively for filters 1 and 2. This is less than 10\% of the
FWHM, so that the effect does not appear obviously in solar images.
Figure~\ref{cakimages} shows a comparison between the CaII K3 Meudon
spectroheliograms (line centre) and MTSP test images (formed at
lower altitude due to the larger bandwidth): filaments and
prominences are no more visible, but sunspots appear more
contrasted; bright faculae look similar, so that the CaII K MTSP
channel produces a good magnetic proxy. It is close to the
wavelength integration of the spectroheliograph data-cube (x, y,
$\lambda$ with 0.093 \AA~ resolution) over the MTSP transmittance.

\section{The two H$\alpha$ channels} \label{sec:ha}

\begin{figure}
   \centering
   \includegraphics[width=\textwidth]{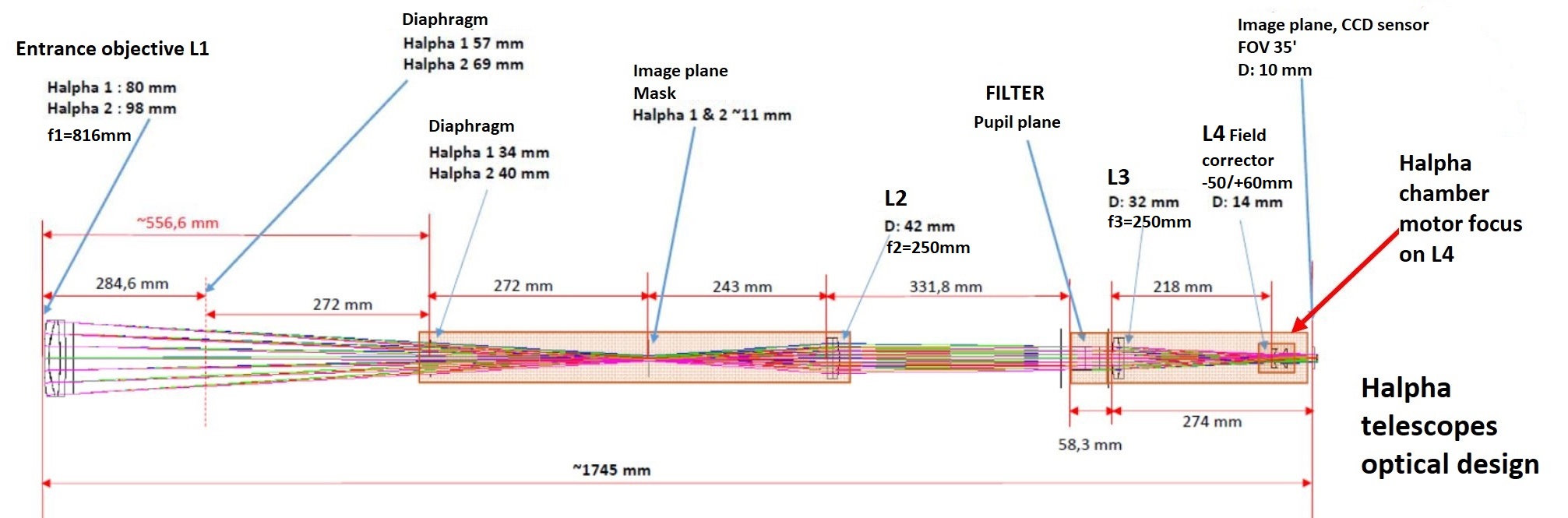}
      \caption{The MTSP H$\alpha$ channels. The filter is located in the pupil
      plane at F/30 inside the L2/L3 afocal system. L4 is a field curvature corrector. The chamber
      has a motor focus (see also Figure~\ref{Telescopes}). The equivalent focal length is 983 mm, as for CaII K.
      The beam
      aperture, in the detector plane, is F/12.3 and F/10.0 respectively for H$\alpha$ 1 and 2.}
         \label{lunetteha}
   \end{figure}

The two H$\alpha$ telescopes (Figure~\ref{lunetteha}) have an
aperture of 80 mm and 100 mm, respectively for filters 1 and 2. They
are composed of a Takahashi TSA102 objective (816 mm focal length)
followed by an afocal chamber (magnification 1.2), including a
DayStar Quantum Pro Fabry-P\'{e}rot filter in the pupil plane. This
system introduces field curvature, so that a field corrector (two
lenses) forms the final image on an interline cooled CCD camera (QSI
660, pixel size = 4.54 $\mu$m = 0.96 arcsec). The Airy spot size is
9.8 and 7.9 $\mu$m, corresponding to 2.06 and 1.65 arcsec
resolution, respectively for telescopes 1 and 2. The system is
protected by an UV/IR cutoff filter in full aperture (not drawn). A
motor focus is integrated to the chambers.

\begin{figure}
   \centering
   \includegraphics[width=\textwidth]{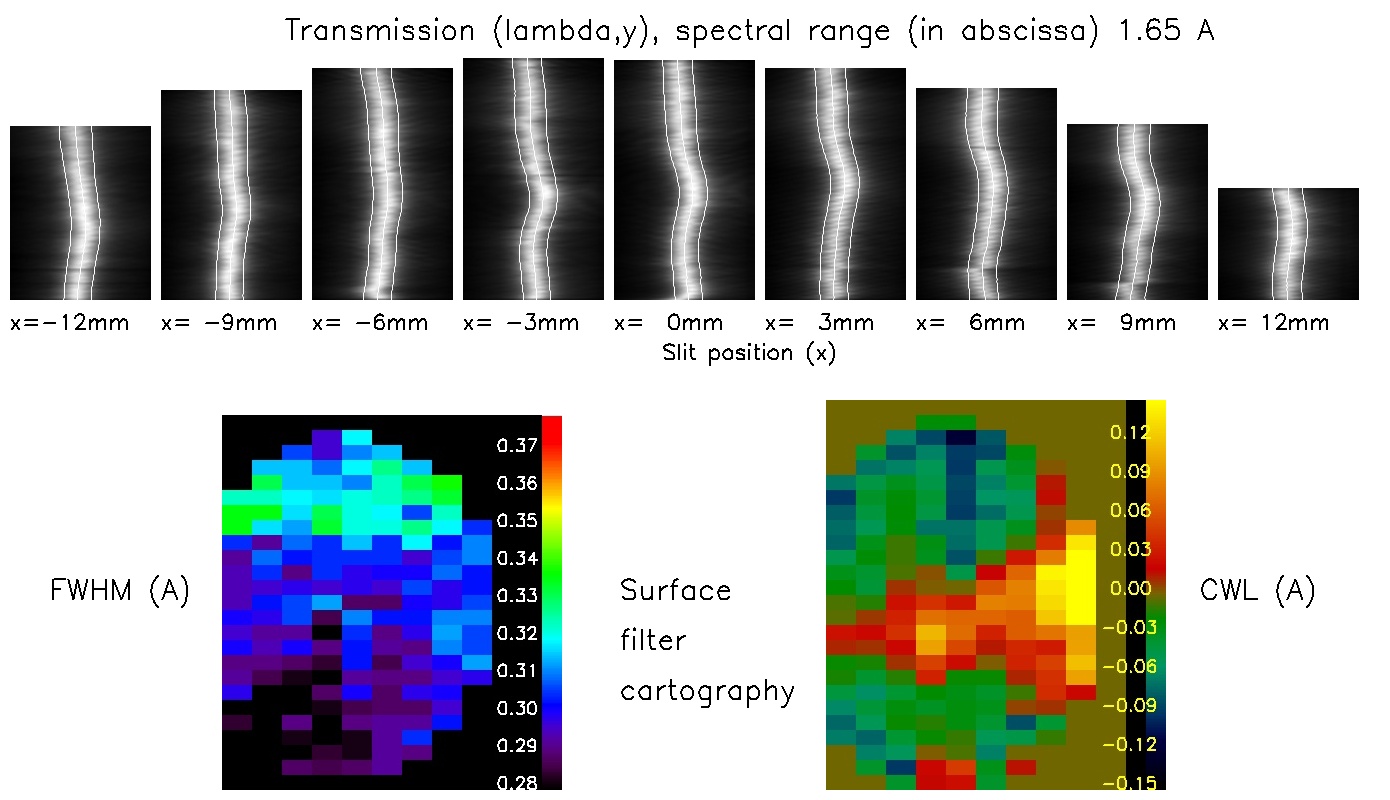}
      \caption{Calibration of H$\alpha$ filter 1. It
      was translated by 3 mm steps in the spectrograph focus of the Meudon Solar Tower (F/75). Top: the H$\alpha$ line
      through the filter for 9 x-positions of the slit (in mm). The white lines delineate the CWL and FWHM (y-direction).
      Bottom: the filter cartography. The local FWHM (\AA) is displayed at left. CWL fluctuations (\AA) are not
      negligible (at right), so that for pupil plane application, the integrated bandpass is enlarged to 0.46 \AA.}
         \label{filterha1}
   \end{figure}

\begin{figure}
   \centering
   \includegraphics[width=\textwidth]{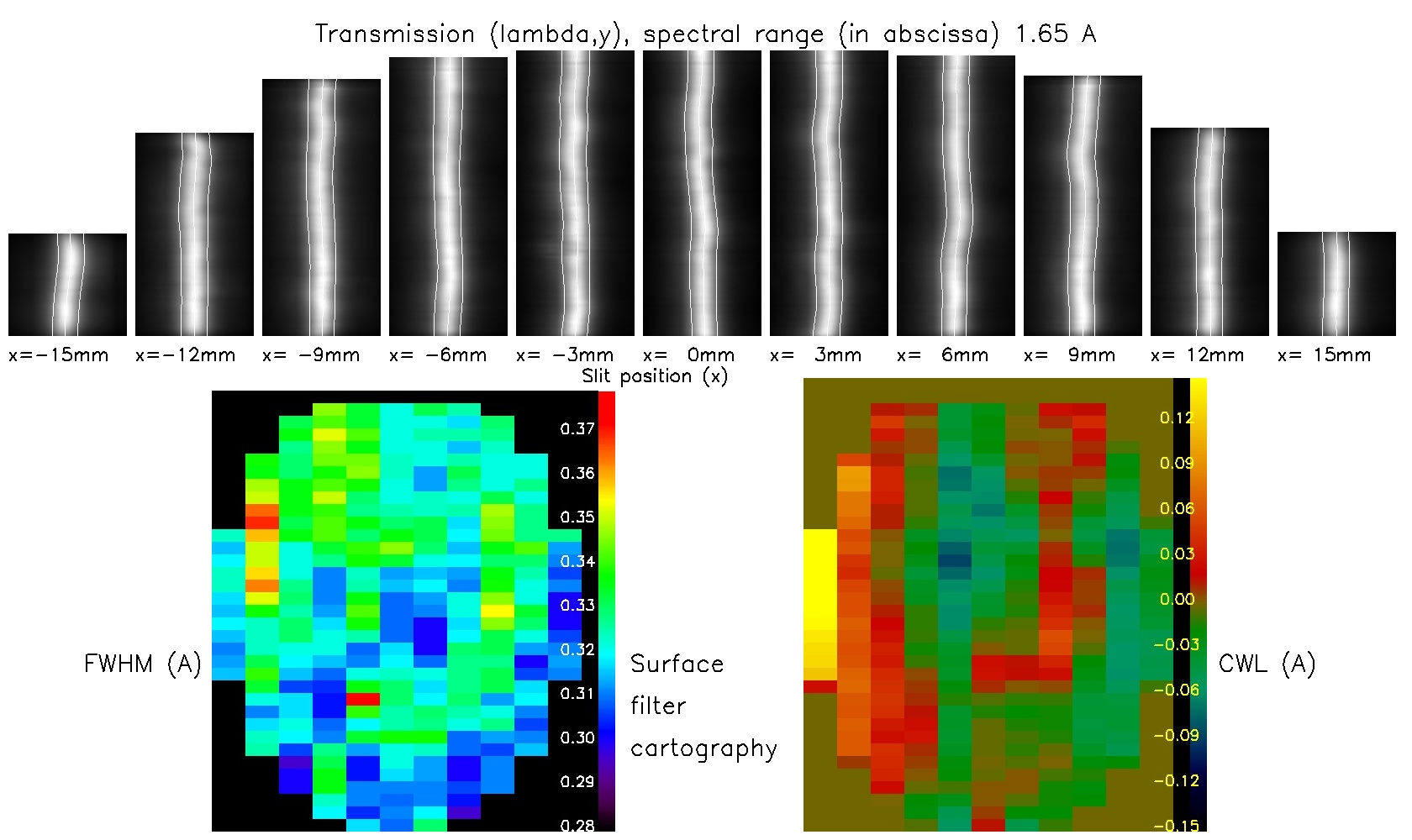}
      \caption{Calibration of H$\alpha$ filter 2. Top: the H$\alpha$ line
      through the filter for 11 x-positions of the spectrograph slit (in mm). The white lines delineate the CWL and FWHM (y-direction).
      Bottom: the filter cartography. The local FWHM (\AA) is displayed at left. At right, the CWL (\AA)
      fluctuations are small, so that the integrated bandpass for pupil plane application (0.34 \AA) remains narrow.}
         \label{filterha2}
   \end{figure}

The calibration of Fabry-P\'{e}rot filters has been done with the
spectrograph of Meudon Solar Tower, at F/75 (much better than the
F/30 filter specification), in order 9 (10 m\AA/pixel). Several
tests have been performed for both filters. First of all, we made a
scan of the surface in order to produce a pupil cartography of the
CWL and FWHM (Figures~\ref{filterha1} and ~\ref{filterha2}). For
that purpose, the filters (31 mm diameter) were translated (-15 mm
$\leq$ x $\leq$ 15 mm) in front of the spectrograph slit by
$\Delta$x = 3 mm steps. ($\lambda$, y) spectral images, with and
without filters, were recorded, for various x-positions (top of
figures). The wavelength transmittance was derived at several (x, y)
locations. The results show that the FWHM of both filters is locally
in the range 0.30-0.35 \AA, but also that filter 2 is better than
filter 1 in terms of CWL uniformity. By integration on the surface,
we computed the resulting bandpass for pupil plane application in
afocal systems, leading to 0.46 and 0.34 \AA, respectively for
filters 1 and 2. The pupil transmittance is Lorentzian shaped in
wavelength and drawn in Figure~\ref{Filters} for both filters.

\begin{figure}
   \centering
   \includegraphics[width=\textwidth]{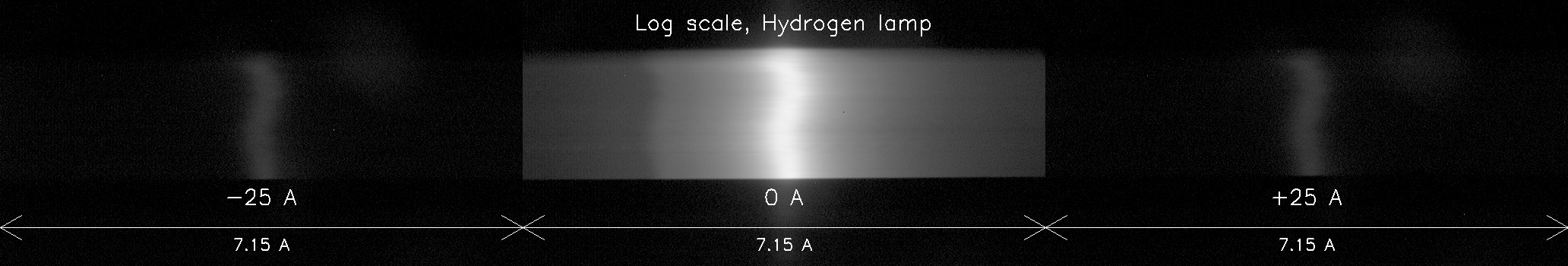}
      \caption{Secondary transmission peaks of H$\alpha$ filter 2 in logarithmic intensity scale (wavelength in abscissa).}
         \label{cannelures}
   \end{figure}

   Fabry-P\'{e}rot filters have secondary lobes, cut by a blocking
   filter. We found, for filter 2, that they appear at $\pm$ 25 \AA~ from the main
   lobe with an intensity of only 0.3\% (Figure~\ref{cannelures}).
   For that filter (0.34 \AA~ FWHM), the measurements provide
   the value of the finesse (75), the reflection coefficient of the cavity (0.96)
   and the interference order for H$\alpha$ line (k = 260).

\begin{figure}
   \centering
   \includegraphics[width=\textwidth]{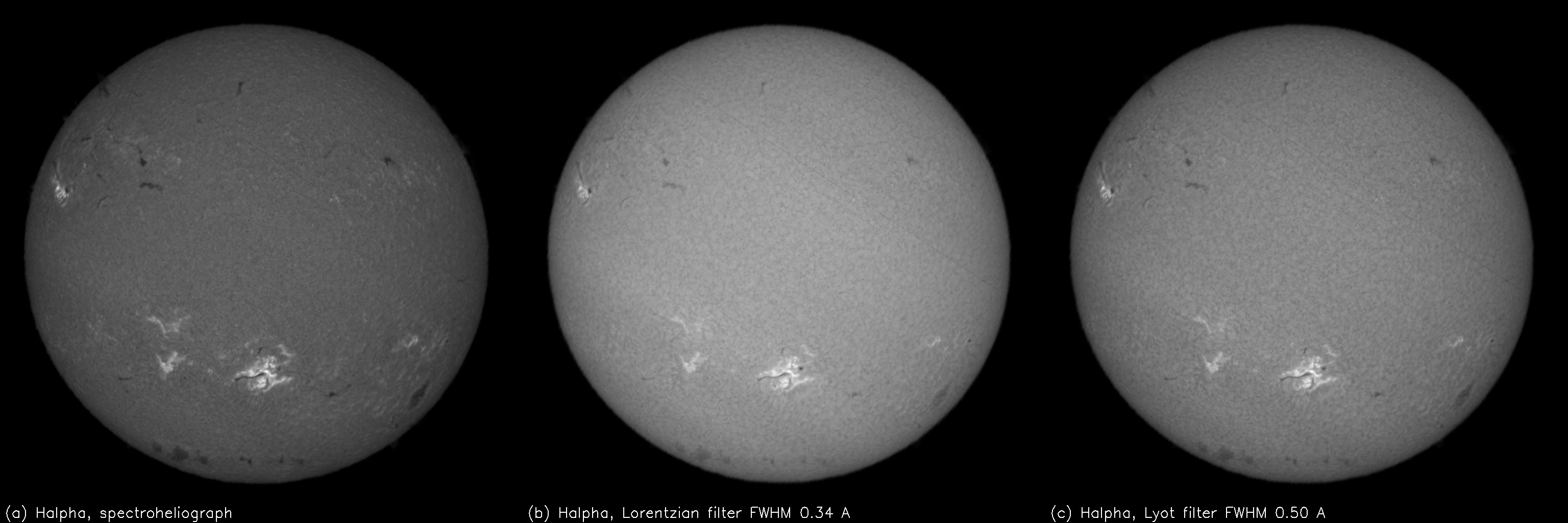}
      \caption{Simulation of H$\alpha$ images obtained after filtering the (x, y, $\lambda$) spectroheliograph data-cube.
      (a) Spectroheliogram at line centre, 28 October 2021.
      (b) Same image, through a 0.34 \AA~ Lorentzian transmission
      (such as MTSP filter 2). (c)  Same image, with a 0.50 \AA~ five stage Lyot filter, for comparison.}
         \label{haimages}
   \end{figure}

Meudon spectroheliograph data, which are made of 3D (x, y,
$\lambda$) data-cubes (0.155 \AA~ wavelength resolution) allow to
simulate images using various transmittances. Figure~\ref{haimages}
displays H$\alpha$ images with different filter characteristics,
such as a Lorentzian transmittance (0.34 \AA~ FWHM) or a 0.50 \AA~
FWHM five stage Lyot transmittance. It clearly shows that a smaller
bandwidth is needed for the Fabry-P\'{e}rot to produce contrasts
similar to the ones provided by Lyot filters. Indeed, the wavelength
curve of Lyot filters drastically cuts the line wings, contrarily to
Lorentzian filters which have extended wings. Hence, the
photospheric light passes in excess and contaminates the contrast of
the chromosphere. When 0.50 \AA~ Lyot filters are sufficient to
select the chromosphere, narrower (0.30 \AA) Fabry-P\'{e}rot devices
(such as MTSP filter 2) are required for the same result.

\begin{figure}
   \centering
   \includegraphics[width=\textwidth]{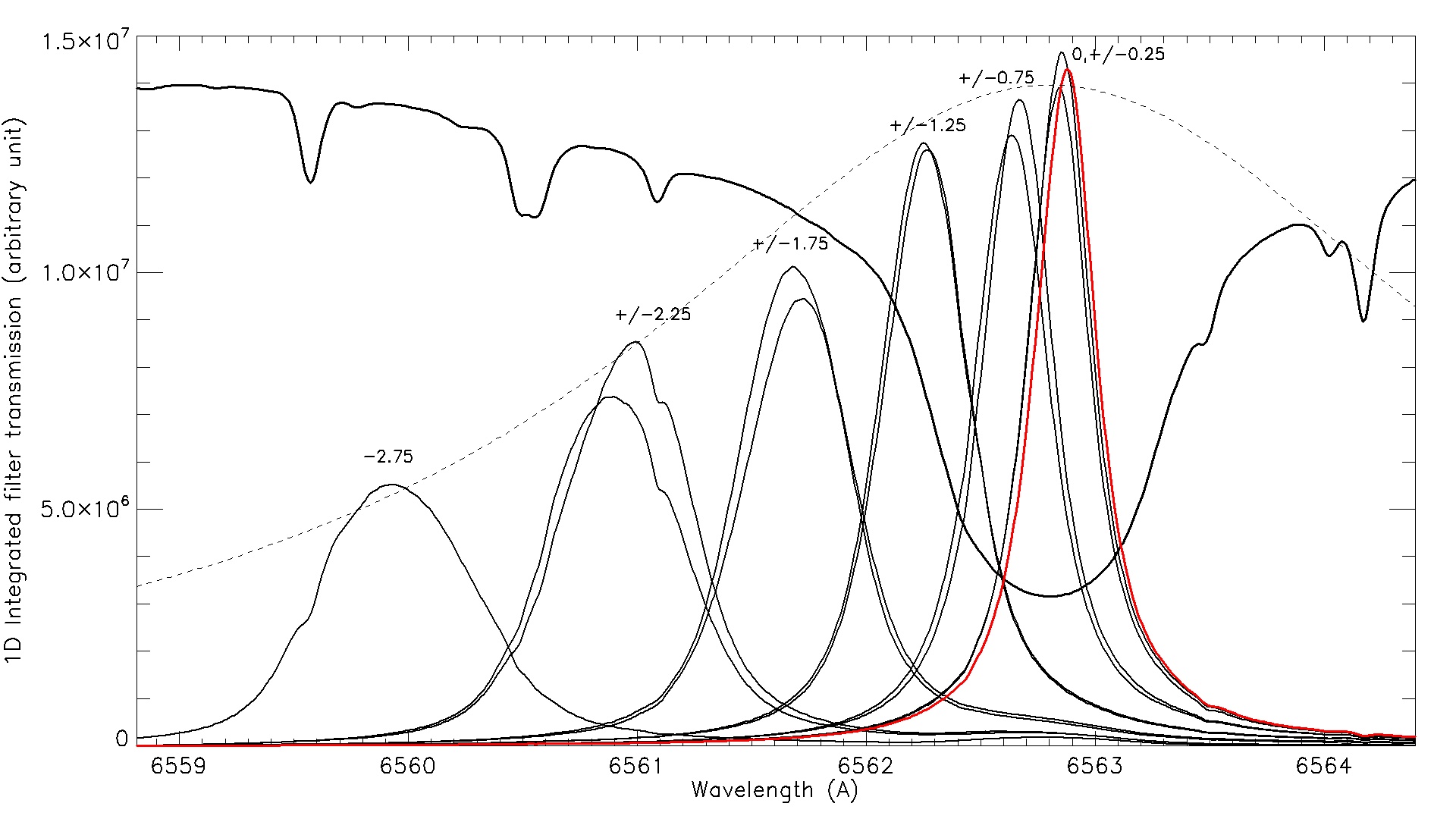}
      \caption{Effect of the tilt angle on the filter transmission curves (abscissa: wavelength in \AA). Tilt values
      are
      0$^{\circ}$ (in red), $\pm$ 0.25$^{\circ}$,
      $\pm$ 0.75$^{\circ}$, $\pm$ 1.25$^{\circ}$, $\pm$ 1.75$^{\circ}$, $\pm$ 2.25$^{\circ}$ and -2.75$^{\circ}$.
      The transmission is blue-shifted with increasing tilt. Thick line:
      the H$\alpha$ profile. Dashed line: the envelope of the blocking filter.}
         \label{tilt}
   \end{figure}

\begin{figure}
   \centering
   \includegraphics[width=\textwidth]{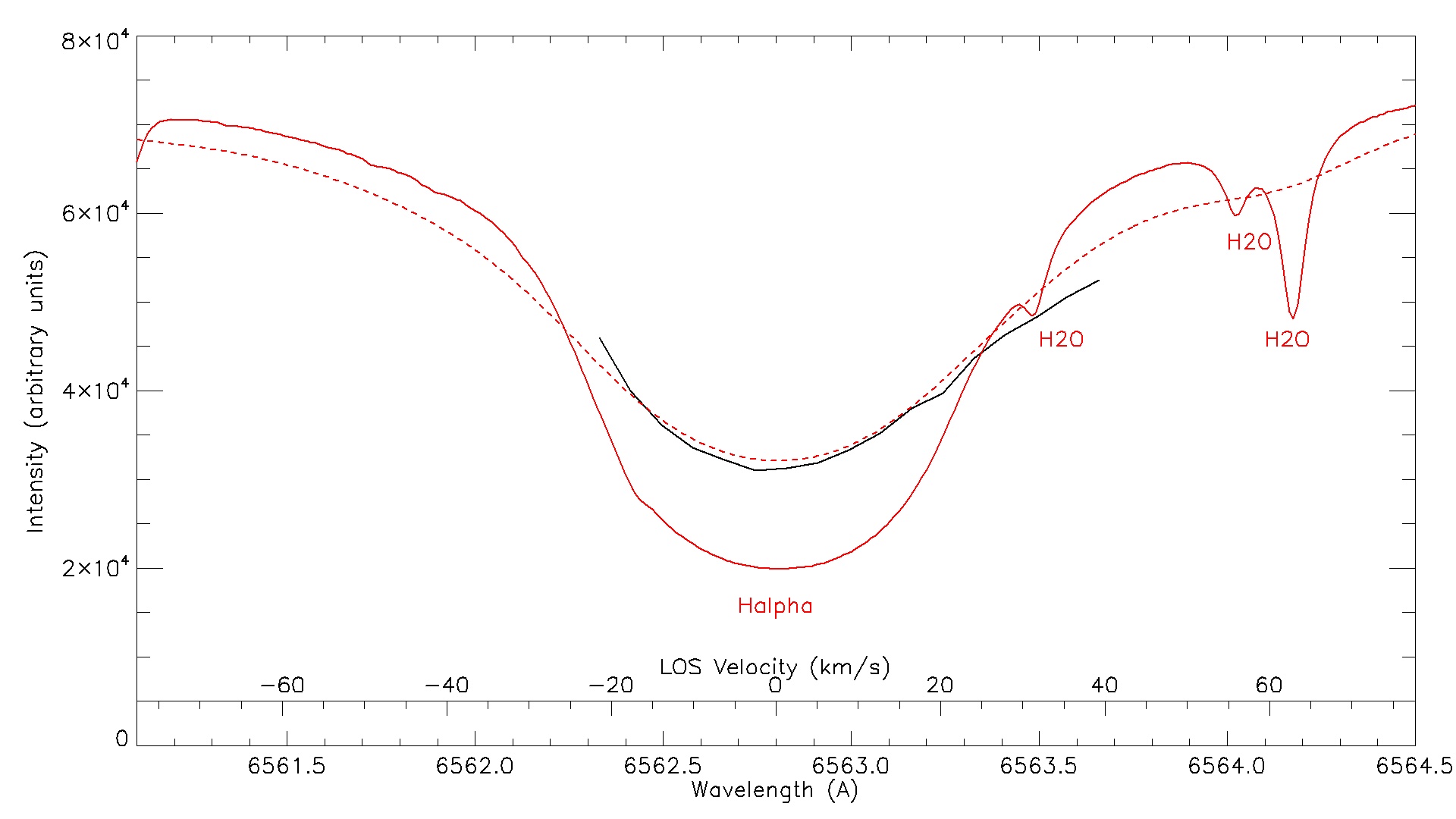}
      \caption{H$\alpha$ line profiles got in imagery or spectroscopic mode.
      Black: MTSP measures resulting from filter temperature variation. Red: the line
      observed with the large spectrograph of Meudon Solar Tower, and the
      convolution by the MTSP filter transmittance (dashed). In abscissa: the
      wavelength (\AA) and the corresponding LOS velocity (km s$^{-1}$).}
         \label{profils}
   \end{figure}

The Solar Tower spectrograph allowed us to explore the angular
dependance of MTSP filters and precise the tilt sensitivity
($\theta$) in terms of CWL and FWHM fluctuations. This experience
also provided an estimate of the effective index of refraction. For
that purpose, the filter was tilted in the range -3$^{\circ}$ $\leq$
$\theta$ $\leq$ +3$^{\circ}$ (Figure~\ref{tilt}). The mean CWL
variations are fitted by the law $\Delta\lambda_{CWL}$ = K
$\theta^{2}$ with K = -0.38 or -0.37 \AA/degree$^{2}$, respectively
for filters 1 and 2. For the tilt dependance of the mean bandpass,
we found FWHM = 0.33 + 0.064 $\theta^{2}$ and 0.35 + 0.052
$\theta^{2}$ (\AA), respectively for filters 1 and 2. This means
that, for a 1$^{\circ}$ tilt, the bandpass is locally enlarged to
0.40 \AA~ and blue-shifted of about -0.40 \AA. But in the afocal
system, at F/30 (half cone angle of 1$^{\circ}$), we have to
integrate the above formulae over angles, which generates a
blue-shift of -0.19 \AA~ (this value was confirmed by the wavelength
line scan made with the filter in imagery mode, which produced
results of Figure~\ref{profils}). However, the F/30 cone angle is
not the only one to consider. The solar diameter is 0.53$^{\circ}$,
which means that the cone incidence $\theta$ varies in the range
-0.26$^{\circ}$ $\leq$ $\theta$ $\leq$ +0.26$^{\circ}$. The
corresponding blue-shift is, at maximum, -0.024 \AA, which can be
neglected, so that the CWL should be almost uniform over the solar
image.

According to the Fabry-P\'{e}rot theory, we have K = - $\frac{1}{2}$
$\frac{\lambda_{0}}{n^{2}}$ (where $\lambda_{0}$ is the line
wavelength). It allows to derive the effective index of refraction n
of the overall filter. We found n $\approx$ 1.62. The distance of
secondary lobes is $\frac{2en}{k^{2}}$ = 25 \AA, where k is the
interference order (260) and e the thickness of the cavity. Hence,
we conclude that e $\approx$ 7.2 mm.

The envelope of transmittance curves of Figure~\ref{tilt}
corresponds to the blocking filter. The estimated FWHM is about 4.5
\AA; this value, considering a Lorentzian shape, is consistent with
the intensity (0.3\%) of the secondary peaks of
Figure~\ref{cannelures}.

\begin{figure}
   \centering
   \includegraphics[width=\textwidth]{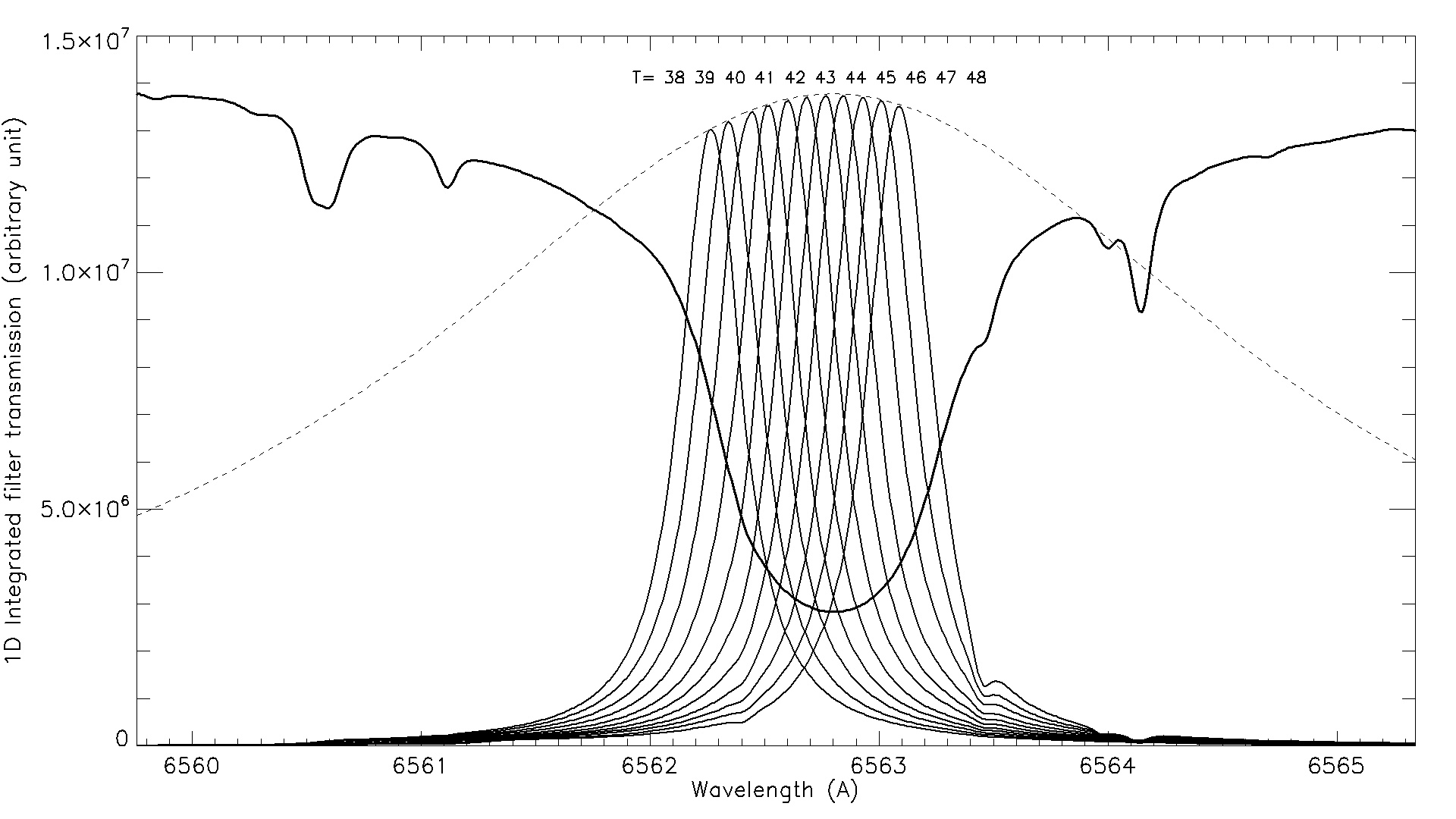}
      \caption{Effect of the temperature on the filter transmission curves (abscissa: wavelength in \AA). Temperature values
      are 38, 39, 40, 41, 41.8, 42.8, 43.8, 44.7, 45.6, 46.6 and 47.5$^{\circ}$C. The transmission is red-shifted with increasing temperature. Thick line:
      the H$\alpha$ profile. Dashed line: the envelope of the blocking filter.}
         \label{temp}
   \end{figure}

The CWL depends on the temperature T. We investigated, by the
spectroscopic means of the Solar Tower, the effect of temperature
changes upon filter 2. The transmittance for temperatures varying
from 38 to 48$^{\circ}$C is reported in Figure~\ref{temp}. The
response is a red-shift corresponding to the linear law
$\Delta\lambda_{CWL}$ = $\kappa$ T - 3.84 \AA, where T is expressed
in $^{\circ}$C and $\kappa$ is the temperature coefficient equal to
0.0874 \AA/$^{\circ}$C. Hence, it is possible to explore the $\pm$
1.0 \AA~ spectral domain centred on the line, but in practice it is
a very slow process. We did not notice any FWHM variation with the
temperature.

The last qualification test was performed with the filter alone, in
image mode, without any spectrograph. Filter 1 was mounted in the
MTSP afocal system, and we explored the wavelength range
6562.0-6563.6 \AA~ by temperature variation. It took a long time,
because the filter needs 10 minutes to stabilize at each wavelength
increment. We recorded also the fluctuations of the solar flux using
a scintillometer, in order to correct intensities measured by the
filter due to atmospheric variations. Then, we derived the H$\alpha$
line profile at the disk centre (Figure~\ref{profils}). We also
plotted the line profile got by spectroscopic means, and convolved
by the Lorentzian transmittance of the filter, and found both
results in good agreement, after correction of the -0.2 \AA~ shift
resulting from the F/30 light cone angle.

\begin{figure}
   \centering
   \includegraphics[width=\textwidth]{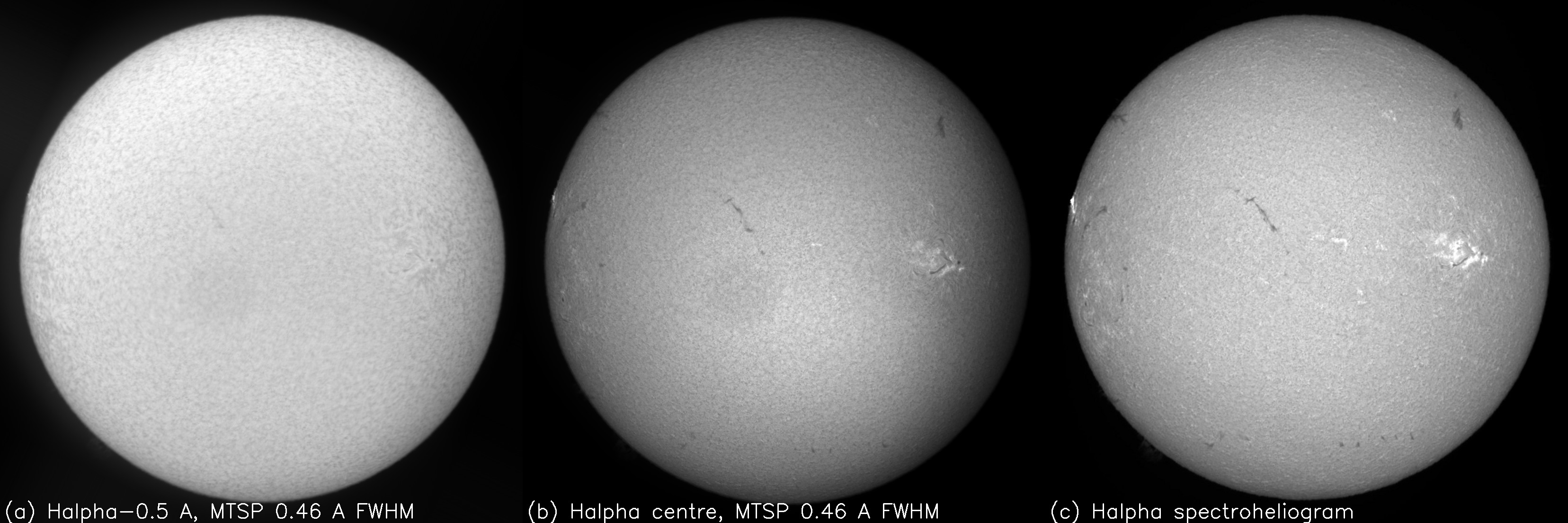}
      \caption{Example of MTSP H$\alpha$ images got during the test campaign of 29 August 2017 with filter 1 (0.46 \AA~ FWHM).
      (a) MTSP, blue wing. (b) MTSP, line centre (0.34 \AA~ FWHM filter
      2 will be used instead in full operation).
      (c) Meudon spectroheliogram for comparison.}
         \label{imgha}
   \end{figure}

Finally, Figure~\ref{imgha} presents the typical H$\alpha$
observations that are scheduled with the above filters. MTSP will
provide almost simultaneously two images, the first one in the line
wing (filter 1, either the blue or the red wing, but not both), and
the second one in the line core (filter 2). We discuss in the next
section the application to the detection of fast evolving
Doppler-shifted events, such as Moreton waves, which require at
least two H$\alpha$ channels and the high observing cadence of 10-15
s.

\section{Moreton waves detection} \label{sec:moreton}

\begin{figure}
   \centering
   \includegraphics[width=\textwidth]{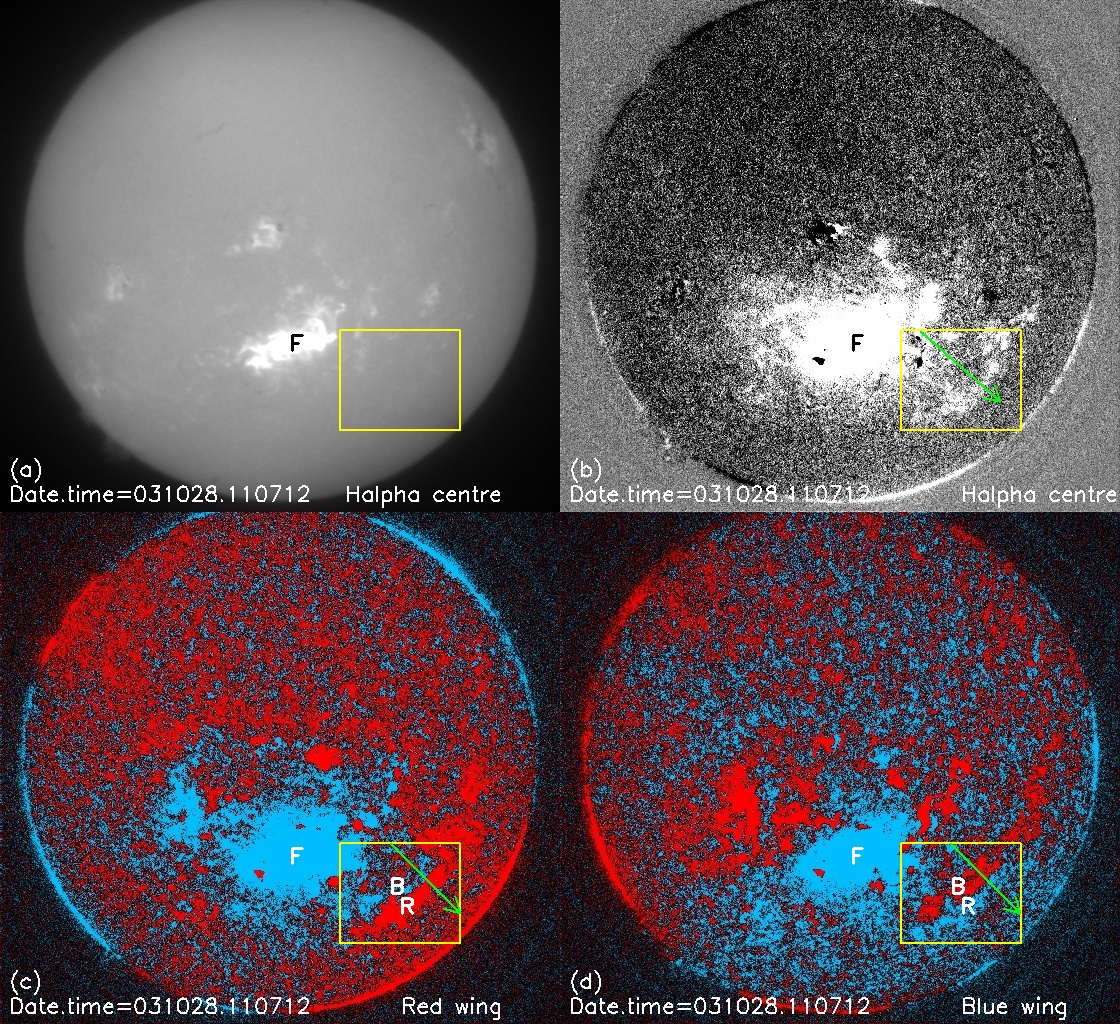}
      \caption{Moreton event of 28 October 2003 at 11:07 UT. Images, in
      H$\alpha$ centre, red and blue wings, were taken with 60 s time step. (a) H$\alpha$ centre.
      (b) H$\alpha$ centre, minus the reference frame at 11:00 UT (event starting time). (c) H$\alpha$ red wing, minus the reference.
      (d) H$\alpha$ blue wing, minus the reference. F, R, B indicate respectively the flare location,
      red and blue shifts of the compression and relaxation fronts of the Moreton wave (propagation direction = green arrow).}
         \label{event1}
   \end{figure}

\begin{figure}
   \centering
   \includegraphics[width=\textwidth]{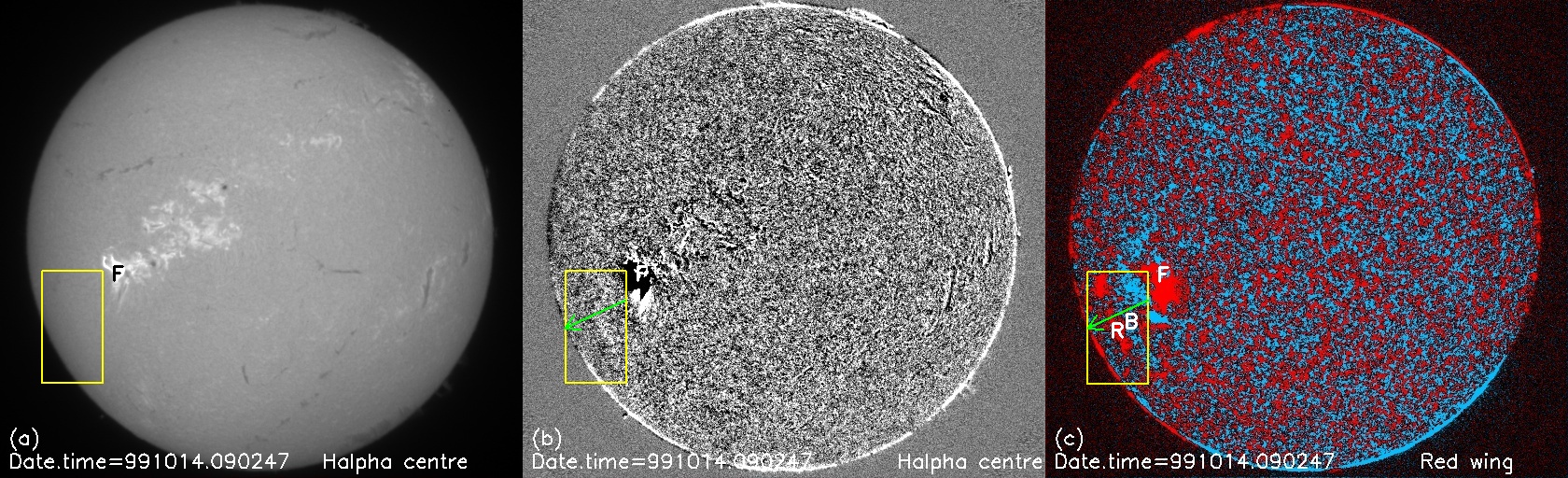}
      \caption{Moreton event of 14 October 1999 at 09:02 UT. Images, in
      H$\alpha$ centre and red wing, were taken with 60 s time step. (a) H$\alpha$ centre.
      (b) H$\alpha$ centre, minus the reference frame at 09:00 UT (event starting time). (c) H$\alpha$ red wing, minus the reference.
      F, R, B indicate respectively the flare location,
      red and blue shifts of the propagating front (propagation direction = green arrow).}
         \label{event2}
   \end{figure}

MTSP is particularly well adapted to the detection of highly dynamic
events, such as Moreton waves originating in energetic flares. As
the filter performances are comparable to those of the previous
Meudon routine (1985-2004), we have chosen two events observed with
this instrument in order to anticipate MTSP capabilities. Moreton
waves appear in H$\alpha$ as fronts propagating at typically 500 km
s$^{-1}$. Such velocities in the chromosphere (8000 K) are so highly
supersonic (C$_{s}$ $\approx$ 10 km s$^{-1}$) and superalv\'{e}nic
(C$_{a}$ $\approx$ 10 km s$^{-1}$ for a 10 G magnetic field) that
they are unlikely of chromospheric nature. Such phenomena last only
a few minutes and are suspected to be the chromospheric counterpart
of coronal waves propagating in the 200 times hotter (1.5 MK) and
more tenuous corona under the form of fast magnetosonic shocks
(C$_{s}$ $\approx$ 150 km$^{-1}$, C$_{a}$ $\approx$ 200 km s$^{-1}$
for a 10 G field). The downward compression of the chromosphere
below the front could be the signature of the coronal shock in
H$\alpha$. Moreton waves occur mainly in X-class flares which are
rare events (about one event/year above X5, six events above X10
since year 2000). We have chosen the famous X17.2 flare of 28
October 2003 (Figure~\ref{event1}), after the solar maximum of cycle
23, and the X1.8 event of 14 October 1999, just before the maximum
(Figure~\ref{event2}). Largest flares often occur during the
descending phase of the solar cycle. In Figure~\ref{event1} (b,c,d
respectively for H$\alpha$ centre and $\pm$0.5 \AA), we have
subtracted the reference frame just before the event at the same
wavelength. Blue/red colours indicate the sign of the resulting
image (blue, or positive, means brighter; red, or negative, means
darker). The wave (yellow box) appears as a brightening in H$\alpha$
core (b). In the red wing subtraction (c), the compression front (R)
is negative (or darker) and corresponds to a red-shift. It is
followed by the relaxation of the chromosphere (B) appearing
positive (or brighter), because blue-shifted. In the blue wing
subtraction (d), this is the contrary: R is positive (or brighter),
because red-shifted, while B is negative (or darker) with the
opposite shift. Figure~\ref{profils} shows that the order of
magnitude of LOS velocities for 0.5 \AA~ shifts is qualitatively 20
km s$^{-1}$. This example shows that Moreton waves can be detected
either by the red or the blue wing. For that reason, MTSP offers
only one wing.

Figure~\ref{event2} presents the X1.8 flare of 14 October 1999,
almost 10 times less energetic than the previous one, so that the
Moreton event is less contrasted. It shows exactly what will provide
MTSP concerning wave detection with one wing. Most events are waited
between 2024 and 2028 around the solar maximum (2025) of cycle 25
(but isolated events, as the X1.0 of 28 October 2021, are possible).
The H$\alpha$ centre (a) will be provided by telescope 2, from which
a reference frame (just before the event) has been subtracted in
(b), showing the brightening front. The Moreton event appears
clearly in the red wing with the red-shifted compression front (R)
followed by the blue-shifted relaxation front (B). The blue wing
could either be chosen. The big difference with the 1985-2004 Meudon
routine is the observing cadence (4 times faster) and the spatial
resolution (2 times better). Hence, MTSP will provide much more
detailed informations to investigate the physics of rare phenomena
at the Sun.

\section{A possible extension for MTSP} \label{sec:exten}

\begin{figure}
   \centering
   \includegraphics[width=\textwidth]{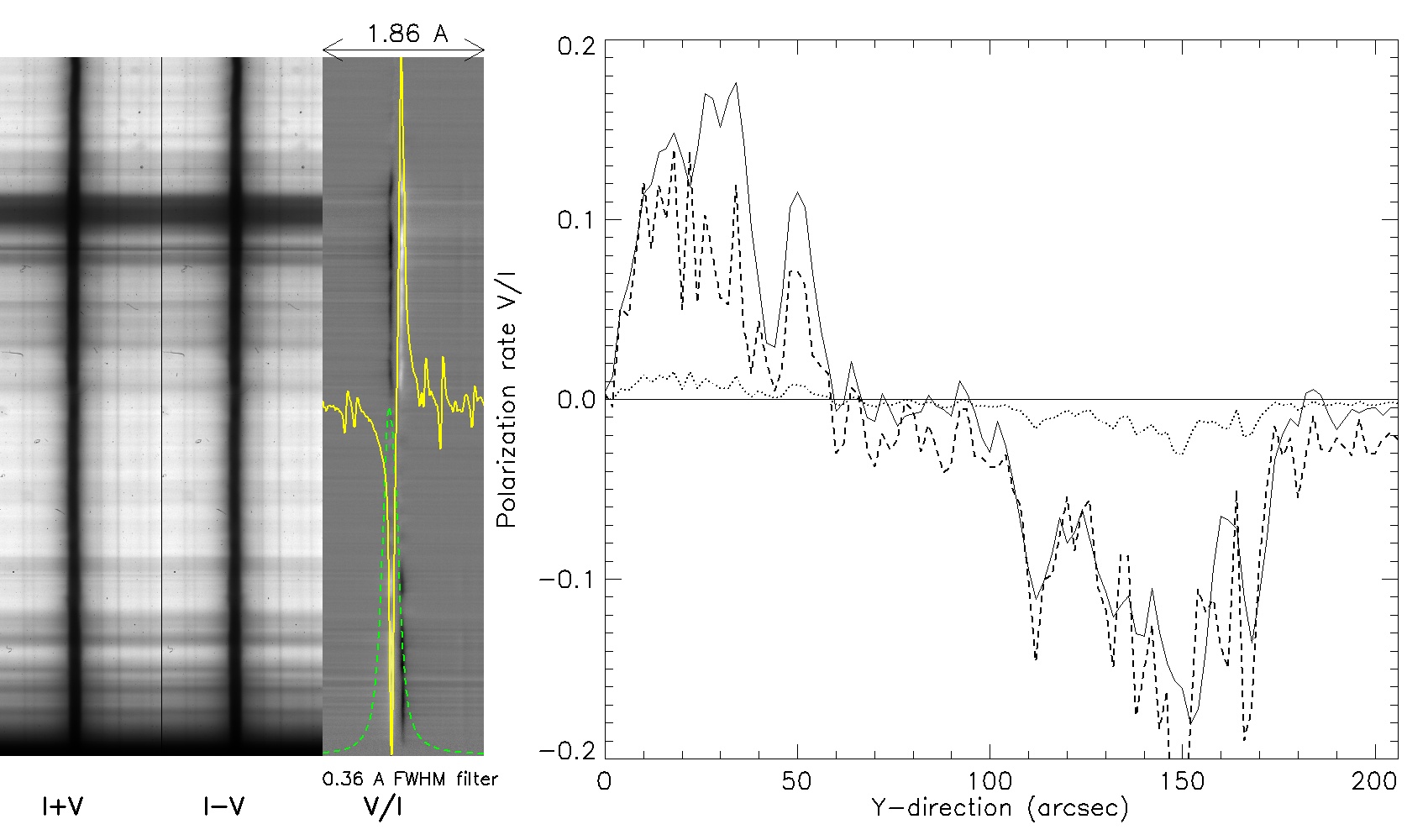}
      \caption{Simulation of polarization measurement with the 0.36 \AA~ FWHM NaD1 filter. Left: spectra of I+V and I-V
      got at Pic du Midi in an active region (wavelength in abscissa, direction of the slit in ordinates, spectral
      pixel 16.8 m\AA). Middle: the polarization
      rate $\frac{V}{I}$ with the quantity $\frac{1}{I}$ $\frac{dI}{d\lambda}$ in the quiet Sun (yellow) and the wavelength
      transmittance of the filter, centred on the blue wing (green, dashed). Right: the polarization rate $\frac{V}{I}$ along
      the slit in the blue wing measured with the spectrograph (solid line) or extrapolated for the filter (dotted line); the dashed line
      is a magnification of the dotted line (ratio given by Figure~\ref{cor}) to
      recover roughly
      the spectroscopic signal. }
         \label{bmag}
   \end{figure}

\begin{figure}
   \centering
   \includegraphics[width=\textwidth]{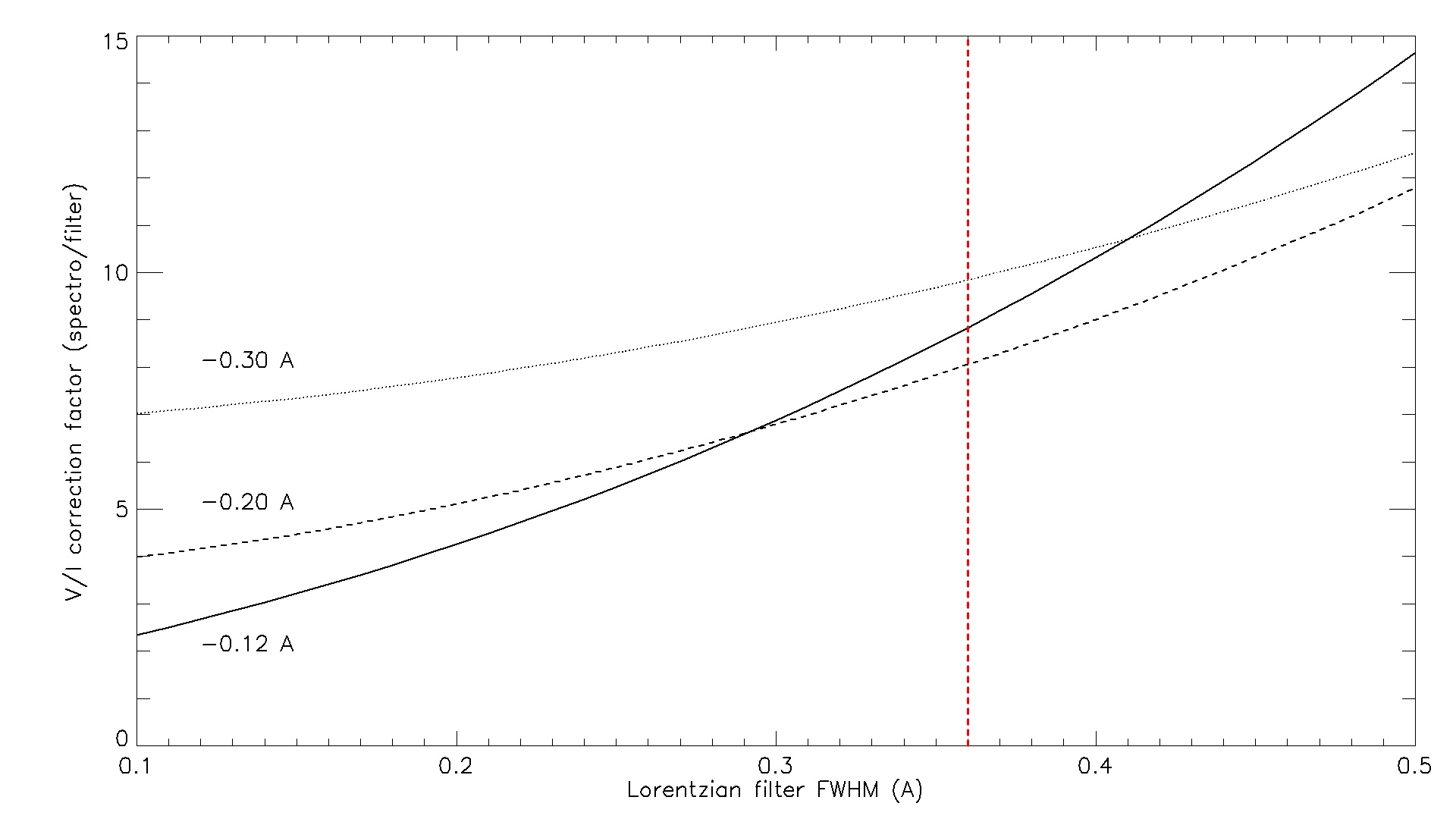}
      \caption{Simulation of the correction factor to apply to the measurements of the polarization rate
      with a Lorentzian filter, as a function of the FWHM (\AA), for various
      CWL shifts (solid, dashed, dotted lines for respectively -0.12, -0.20, -0.30 \AA~ CWL shifts).
      The blue inflexion point of the line is located at -0.12 \AA. The red
      (dashed) line indicates our filter FWHM; the best CWL shift is -0.20 \AA. }
         \label{cor}
   \end{figure}

We own a NaD1 5896 \AA~ Fabry-P\'{e}rot filter manufactured by
DayStar (0.36 \AA~ FWHM), for mounting in an afocal design at F/30
similar to the one of the H$\alpha$ telescopes described above.
Magnetograms of the Sun have been successfully produced in this line
(Land\'{e} factor 1.33) by Mount Wilson (full disk) or by the
Narrow-band Filter Imager (NFI) onboard HINODE for small regions.
NaD1 is a Fraunhofer line formed in the low chromosphere above the
FeI 6173 \AA~ line observed in the photosphere by the Helioseismic
and Magnetic Imager (HMI) onboard SDO. As DayStar filters are
linearly polarizing (made of a birefringent material), the
incorporation of a Liquid Crystal Variable Retarder (LCVR) at the
primary focus, providing $\pm \frac{\lambda}{4}$ fast modulation
\citep{Malherbe2007} suffice to produce alternatively I+V and I-V
images (I, V for Stokes parameters). From the circular polarization
rate $\frac{V}{I}$ and the weak field theory (see
\cite{Stenflo1994}), it is possible to estimate the LOS magnetic
field (BLOS) together with the field polarity. Indeed, $\frac{V}{I}$
is proportional to BLOS and $\frac{1}{I}$ $\frac{dI}{d\lambda}$.
This quantity depends on the wavelength and is maximum when measured
at the inflexion points of the line. Figure~\ref{bmag} displays
spectroscopic observations obtained with this method at the 8 m
spectrograph of the Pic du Midi Turret Dome. We have numerically at
the inflexion points $\frac{V}{I} \approx 2.2 10^{-4} BLOS$, where
BLOS is expressed in Gauss. As the slope of NaD1 wings is steep,
Stokes V profiles are sharp. We observed polarization rates up to
0.20 (BLOS $\approx$ 1000 G) in some sunspots. In order to estimate
the measurement capabilities in imagery, we have integrated the line
profiles over the wavelength transmission of the NaD1 filter. The
$\frac{V}{I}$ signal becomes much smaller, because the filter FWHM
($W_{1}$)  is much larger than the half width ($W_{2}$) of the line
derivative $\frac{dI}{d\lambda}$, which concentrates the
polarimetric signal in a very narrow wave-band (Figure~\ref{bmag}).
Hence, the corrective factor to apply to the polarization rate is
roughly equal to the $\frac{W_{1}}{W_{2}}$ ratio and plotted in
Figure~\ref{cor} for various CWL and $W_{1}$ values. Despite of the
signal loss due to wavelength integration, it is of interest to
consider polarimetric measurements in imagery. For that purpose, a
NaD1 telescope will be tested at Meudon in the coming year in the
context of a possible extension for MTSP. As the polarization rate
to measure with the filter has the same order of magnitude than the
photon noise of a single exposure (1\% or 400 G), it is necessary to
acquire many frames in order to select best images and reduce the
noise, as done by \cite{Roudier2006}. For example, 100 couples (I+V,
I-V) will decrease the noise to 40 G. 20 G should be achieved with
either 2 $\times$ 2 binning or with 400 couples. It must be noticed
that a fast observing cadence is not required for LOS magnetograms,
because the magnetic field evolves on longer time scales. The method
is also, in principle, valid for the H$\alpha$ telescopes, but in
practice, H$\alpha$ is a broad line and the sensitivity to the
magnetic field is reduced by the factor 6 in comparison to NaD1.

\section{Conclusion} \label{sec:Conclu}

The MTSP project is dedicated to the survey of fast evolving events
in the chromosphere at the source of solar activity, such as flares
and coronal mass ejections. Large flares often occur after the solar
maximum (2025 for the present cycle) during a few years. MTSP, with
an outstanding cadence of 10-15 s, has also the major goal to
investigate Moreton waves, which are extremely fast and rare
phenomena associated with largest flares. Such events are difficult
to detect in the chromosphere, so that only a few cases have been
studied. With systematic, fast and multi-channel observations (two
H$\alpha$ and one CaII K telescopes), MTSP will increase the chances
to catch such phenomena. Data could be combined to SDO/AIA
observations of the low corona at slower cadence (45 s) in several
EUV channels. MTSP will operate automatically at Calern observatory
(1270 m) under good seeing and climatic conditions. High cadence
observations will be freely delivered, without any delay, to the
international community through a dedicated database located at Nice
computer centre. MTSP will cover cycle 25, from 2023 to, at least,
the end of the present decade, where new generation solar synoptic
networks could start, such as the Next Generation GONG project
\citep{Hammel2019} or the Solar Physics Research Integrated Network
Group (SPRING, \cite{Gosain2018}). MSTP will support Solar Orbiter
(ESA) and Parker Solar Probe (NASA) operations in the coming years.

\section{Disclosure of potential conflicts of interest} \label{sec:Discl}

The authors declare that they have no conflicts of interest.

\section{Data availability statement} \label{sec:Discl}

The authors declare that the datasets analysed during the current
study are publicly available from the corresponding author upon
request.

\begin{acknowledgements}

We thank the anonymous referee for useful suggestions and comments.
We are indebted to Y. Bresson, C. Renaud (OCA), J.-M. Rees, C.
Blanchard (OP) and the technical teams of OP and OCA for their
assistance. We are also grateful for financial support to Paris and
Nice observatories, the Direction G\'{e}n\'{e}rale de l'Armement,
the IDEX Plas$@$Par, the Programme National Soleil
Terre (PNST/INSU/CNRS), the DIM ACAV (Ile de France Region) and the IDEX UCA/JEDI.\\

\end{acknowledgements}

\bibliographystyle{spbasic}      % basic style, author-year citations

\bibliography{papier}

\end{document}